\documentclass[aps,prd,11pt,twoside,tightenlines,nofootinbib,showpacs,preprint,superscriptaddress]{revtex4-1}
\usepackage[colorlinks=true]{hyperref}
\usepackage{xcolor}
\hypersetup{colorlinks=true, citecolor=blue, urlcolor=blue, linkcolor=blue}
\usepackage{amsmath,amssymb}
\usepackage[final]{graphicx}
\usepackage{subcaption}
\usepackage{float}

\begin{document}

\title{Study of $\bar{B}^0_{(s)}$ decaying into $D^0$ and  $\pi^{+}\pi^{-}$, $K^{+}K^{-}$, or $\pi^{0}\eta$ with final state interactions}

\author{Hiwa A. Ahmed}
\email{hiwa.ahmed@charmouniversity.org}
\affiliation{Medical Physics Department, College of Medicals and Applied Science, Charmo University, 46023, Chamchamal, Sulaimania, Iraq}
\affiliation{School of Nuclear Science and Technology, University of Chinese Academy of Sciences, Beijing 100049, China}

\author{C. W. Xiao } 
\email{xiaochw@csu.edu.cn}
\affiliation{School of Physics and Electronics, Hunan Key Laboratory of Nanophotonics and Devices, Central South University, Changsha 410083, China} 

\date{\today}

\begin{abstract}

We investigate the resonant contributions in the decays $\bar{B}^0_{(s)} \to D^0 \pi^{+}\pi^{-}$, $\bar{B}^0_{(s)} \to D^0 K^{+}K^{-}$, and $\bar{B}^{0} \rightarrow D^{0} \pi^{0}\eta$ by considering the final state interactions within the chiral unitary approach. The scalar resonances produced dynamically from the final state interactions of these decays are the ones of $f_{0}(500)$, $f_{0}(980)$, and $a_{0}(980)$. In addition, the vector mesons $\rho$ and $\phi$ produced in the $p$-wave are also taken into account. From the invariant mass distributions of $B^{0}$ decays, one can see that the contributions of the states $f_{0}(500)$ and $a_{0}(980)$ are remarkably larger than the one of $f_{0}(980)$. However, for the case of $B^{0}_{s}$ decays only a clear structure closed to the $K^{+} K^{-}$ threshold appears, which corresponds to the $f_{0}(980)$ state and has no signal for the $f_{0}(500)$ resonance. The dominant contributions are from the vector meson $\phi$ for the decay $\bar{B}^0_s \to D^0 K^{+}K^{-}$, where the results of the $K^{+} K^{-}$ invariant mass distributions are in agreement with the experimental data up to $1.1$ GeV. Moreover, the branching fractions and the ratios of branching fractions are also investigated, where some of them are consistent with the experimental measurements, and the branching fraction $\text{Br}(\bar{B}^{0} \rightarrow D^{0} a_{0}(980) [a_{0}(980) \to \pi^{0}\eta])=(6.08 \pm 0.72)\times 10^{-5}$ and the ratio of branching fractions $\text{Br}(\bar{B}^{0}\rightarrow D^0 f_{0}(980))/\text{Br}(\bar{B}^{0}\rightarrow D^0 a_{0}(980))=0.13 \pm 0.03$ are predicted.

\end{abstract}
\pacs{}

\maketitle

\section{Introduction}

For investigating the $CP$-violation phenomena and understanding the physics beyond the Standard Model (SM), many non-leptonic three-body $B$ meson decays have been measured by the collaborations Belle, BaBar, CLEO, LHCb, and so on. 
The charmless hadronic $B$ decays with three-pseudoscalar final states were observed by the CLEO collaboration \cite{Eckhart:2002qr}, where the branching fractions of the decays $\bar{B} \rightarrow K^{0} \pi^{+} \pi^{-}$ and $\bar{B} \rightarrow K^{*+}(892) \pi^{-}$ were measured. 
In Ref. \cite{Lees:2012kxa}, the BABAR collaboration studied the decays $B^{0} \rightarrow K^{+} K^{-} K_{s}^{0}$, $B^{+} \rightarrow K^{+} K^{-} K^{+}$, and $B^{+} \rightarrow K_{s}^{0} K_{s}^{0} K^{+} $ with the amplitude analysis of Dalitz plot approach, and measured the $CP$-violation parameters based on the collected data sample of approximately $470 \times 10^{6} $  $B \bar{B}$ decays. Moreover, the Belle collaboration measured the $CP$-violation asymmetries in the $B^{0} (\bar{B}^{0}) \rightarrow K^{+} K^{-} K_{s}^{0}$ decays using the time-dependent Dalitz approach \cite{Nakahama:2010nj}, where the direct $CP$-violation signal was not found. 
Recently, the LHCb collaboration has reported lots of the results on the investigation of the $CP$-violation phenomena. The decays $\bar{B}_{s}^{0} \rightarrow J/\psi f_{0}(980)$ and $\bar{B}_{s}^{0} \rightarrow \phi \pi^{+}\pi^{-}$ were observed for the first time by the LHCb collaboration in Ref. \cite{Aaij:2011fx} and Ref. \cite{Aaij:2016qnm}, respectively, which could be used to detect the $CP$-violation in the $B_{s}^{0}$ decays. The amplitude analysis was performed for the decays $B_{s}^{0} \rightarrow K_{s}^{0} K^{\pm}\pi^{\mp}$, $\bar{B}^{+} \rightarrow \pi^{+} \pi^{+}\pi^{-}$, and $\bar{B}^{+} \rightarrow D^{+} D^{-} K^{+}$ in Refs. \cite{Aaij:2019nmr,Aaij:2019jaq,Aaij:2020ypa}, where the $CP$-violation effect was discussed and the resonant contributions were found. In Ref. \cite{Aaij:2017zgz}, the resonance structures were analyzed in details for the $\bar{B}_{s}^{0} \rightarrow J/\psi K^{+} K^{-}$ decay with the $CP$-violation phase measured. In addition, to hint the $CP$-violation effect, the relative branching ratios for the decays $B^{+} \rightarrow K^{+} K^{+} K^{-}$, $\pi^{+} K^{+} K^{-}$, $K^{+} \pi^{+} \pi^{-}$, and $\pi^{+} \pi^{+} \pi^{-}$, were measured precisely in Ref. \cite{Aaij:2020dsq} with a model-independent approach. 
The $\bar{B}^{0} \rightarrow D^{0}\pi^{+}\pi^{-}$ decay was investigated by the Belle collaboration \cite{Kuzmin:2006mw}, and the total branching fraction of the three-body decay $ Br(\bar{B}^{0} \rightarrow D^{0}\pi^{+}\pi^{-}) = (8.4 \pm 0.4(stat.)\pm 0.8(syst.))\times 10^{-4} $ was reported, where the intermediate resonance contributions were taken into account. The LHCb collaboration measured the $\bar{B}^{0} \rightarrow D^{0}\pi^{+}\pi^{-}$ decay more precisely with the Dalitz plot technique and large data sample \cite{Aaij:2015sqa}, of which the measured branching ratio was $(8.46 \pm 0.14 \pm 0.29 \pm 0.40) \times 10^{-4}$, and where many resonances contributions were found. 
In Ref. \cite{Aaij:2015rqa}, the $\bar{B}_{s}^{0} \rightarrow D^{0} f_{0}(980)$ decay was searched by LHCb collaboration using the collected data of $p p $ collisions during 2011 and 2012, where no significant signal was found and the upper limits on its branching fraction were set. Moreover, the LHCb collaboration reported the first observations of the decays $\bar{B}_{(s)}^{0} \rightarrow D^{0} \phi$ \cite{Aaij:2018jqv} and $\bar{B}_{(s)}^{0} \rightarrow D^{0} K^{+} K^{-}$ \cite{Aaij:2018rol} with the corresponding branching fractions determined. 

Except for the hint of $CP$-violation, the two-body resonance information in the three-body $B_{(s)}$ decays is also caught much theoretical attention.
Utilizing the light-cone sum rule approach, Ref. \cite{Shi:2021ste} investigated the $B^- \to K^+ K^- \pi^-$ decay with the contributions from the scalar, vector and tensor resonances. 
Employing the factorization-assisted topological-amplitude approach, the three-body charmed $B_{(s)}$ decays were studied in Ref. \cite{Zhou:2021yys} by considering the resonant effects, where the related branching fractions and the SU(3) symmetry breaking effect were calculated. 
With the perturbative QCD (pQCD) approach, the charmless decays $B_{(s)} \to V \pi\pi$ ($V$ presented the vector meson) were studied in Ref.  \cite{Yang:2021zcx}, where the $f_0(980)$ resonance contribution was introduced in the calculation of the corresponding branching fractions. 
The $f_0(980)$-$f_0(500)$ mixing was discussed in Ref. \cite{Zhang:2016qvq} for the charmed $B_{(s)}$ decays under the pQCD approach.
The decays $\bar{B}_{(s)}^{0} \rightarrow D^{0}(\bar{D}^{0}) \pi^{+} \pi^{-}$ were studied in Ref. \cite{Xing:2019xti} within the pQCD approach, where only the $s$-wave contributions were taken into account and the Breit-Wigner model was used for the scalar form factors of the resonances $f_{0}(500)$, $f_{0}(980)$, $f_{0}(1500)$ and $f_{0}(1700)$. 
Also using the pQCD formalism, Ref. \cite{Li:2020zng} investigated the three-body decays $B_{(s)} \rightarrow [D^{(*)}, \bar{D}^{(*)}] K^{+} K^{-}$ by considering the resonance contributions in the $s$-, $p$-, and $d$-waves. 
Analogously, applying the pQCD approach, the resonance contributions were taken into account in the $B_{(s)}$ three-body decays in Refs. \cite{Zou:2020atb,Ma:2020jsb,Li:2021cnd}. 
Based on the QCD factorization approach and constrained by the analyticity and unitarity, the parametrizations of three-body $B$ and $D$ weak decay amplitudes were discussed in detail in Ref. \cite{Boito:2017jav} with the isobar model for the quasi-two-body resonance contributions. 
In Ref. \cite{Liang:2014ama}, the decays of $\bar{B}_{(s)}^{0}$ into a $D^{0}$ meson and a scalar meson or a vector meson were researched using the final state interaction approach, where the scalar resonances $f_{0}(500)$, $f_{0}(980)$, and $a_{0}(980)$, were dynamical generated in the subprocesses of quasi-two-body interaction with the chiral unitary approach (ChUA) \cite{Oller:1997ti,Oset:1997it,Kaiser:1998fi,Oller:2000fj,Oller:2000ma,Hyodo:2011ur}. Furthermore, also combined with the final state interaction approach and the ChUA, three-body hadronic $B_{(s)}$ decays were analyzed in Refs. \cite{Liang:2014tia,Liang:2015qva,Bayar:2014qha} to study the dynamical production properties of the scalar resonances $f_{0}(500)$, $f_{0}(980)$ and $a_{0}(980)$. More discussions and applications about this approach can be referred to Ref. \cite{Oset:2016lyh}.

Recently, based on the BESIII measurements, Refs. \cite{Achasov:2020qfx,Achasov:2020aun,Achasov:2021dvt} discussed the four-quark nature of the states $f_{0}(500)$, $f_{0}(980)$ and $a_{0}(980)$ in the $D_{(s)}$ semileptonic decays and the $J/\psi$ radiative decays, where the four-quark nature of these resonances had been early concerns in Refs.~\cite{Achasov:1979xc,Achasov:1980gu,Achasov:1980tb,Achasov:1981kh,Achasov:1987ts,Achasov:2003cn}. Also based on the BESIII measurements and with the final state interaction under the ChUA, Ref. \cite{Molina:2019udw} investigated the so-called pure $W$-annihilation decay $D^+_s \to \pi^+ \pi^0 \eta$, and Ref. \cite{Duan:2020vye} studied the single Cabibbo suppressed process $D^+ \rightarrow \pi ^{+} \pi ^{0} \eta$, where the main role of the $a_{0}(980)$ state generated in the subprocesses was found. In Ref. \cite{Ikeno:2021kzf} the production of the $a_0(980)$ resonance was discussed in detail for the single Cabibbo suppressed decays $D^+ \to \pi^+ \eta \eta$ and $D^+ \to \pi^+ \pi^0 \eta$, where both the internal and external $W$-emission mechanisms were taken into account. Moreover, the reactions $D^+\to K^-K^+K^+$ and $D_s^+ \to \pi^+ K_s^0 K_s^0$ were investigated in Refs. \cite{Roca:2020lyi} and \cite{Dai:2021owu}, respectively, where some resonances were generated in the two-body final state interactions. The decays $B^{0}_{(s)} \to \phi \pi^+\pi^-$ were researched in our former work of Ref. \cite{Ahmed:2020qkv}, where the productions of the $f_{0}(980)$ resonance and some vector mesons were examined. Thus, in the present work, inspired by the new results of the LHCb collaboration for the decays $\bar{B}_{(s)}^{0} \rightarrow D^{0} \phi$ \cite{Aaij:2018jqv} and $\bar{B}_{(s)}^{0} \rightarrow D^{0} K^{+} K^{-}$ \cite{Aaij:2018rol} as discussed before, we continue to investigate the $\bar{B}^{0}_{(s)}$ decays with the processes of decaying into a $D^{0}$ meson and a pair of pseudoscalar mesons $\pi^{+}\pi^{-}$, $\pi^{0} \eta$ and $K^{+} K^{-}$ within the final state interaction approach, aiming to study the productions of the resonances $f_{0}(500)$, $f_{0}(980)$ and $a_{0}(980)$ in the three-body decay processes. 

Our work is organized as following. In the next section, we will introduce the formalism of the final state interaction in $s$-wave within the ChUA. Furthermore, we also discuss the vector meson productions in $p$-wave under the same weak decay mechanism. In the following section, we show the results of the invariant mass distributions and the branching fractions of corresponding decay channels. At the end, we make a short conclusion.

\section{Formalism}  

In the present work, we investigate the decays $\bar{B}^{0}_{(s)} \rightarrow D^{0}\; \pi^{+}\pi^{-}( \pi^{0} \eta, K^{+} K^{-})$ within the final state interaction approach, and take into account the $s$- and $p$-waves resonance contributions, which are from the resonances $f_{0}(500)$, $f_{0}(980)$, and $a_{0}(980)$ in the $s$-wave productions, and the states $\rho$, $\omega$, and $\phi$ in the $p$-wave case. 
In our formalism for the hadroniazation mechanism of the three particles produced in the final states of the decay processes $\bar{B}^{0}_{(s)} \rightarrow D^{0}\; \pi^{+}\pi^{-}(K^{+} K^{-}, \pi^{0} \eta)$, at the quark level the bottom quark goes through the weak transitions $b \to c \bar{u} d$ and $b \to c \bar{u} s$ for the $\bar{B}^{0}$ and $\bar{B}^{0}_{s}$, respectively, where the $c \bar{u}$ quarks produce the $D^{0}$ meson and the remaining pair of quarks undergoes the hadronization procedure to generate the meson pairs $\pi^{+}\pi^{-}$, $\pi^{0}\eta$, or $K^{+} K^{-}$ in the final states; see more details below. 

\begin{figure}[H]
\begin{subfigure}{0.49\textwidth}
  \centering
  \includegraphics[width=1\linewidth]{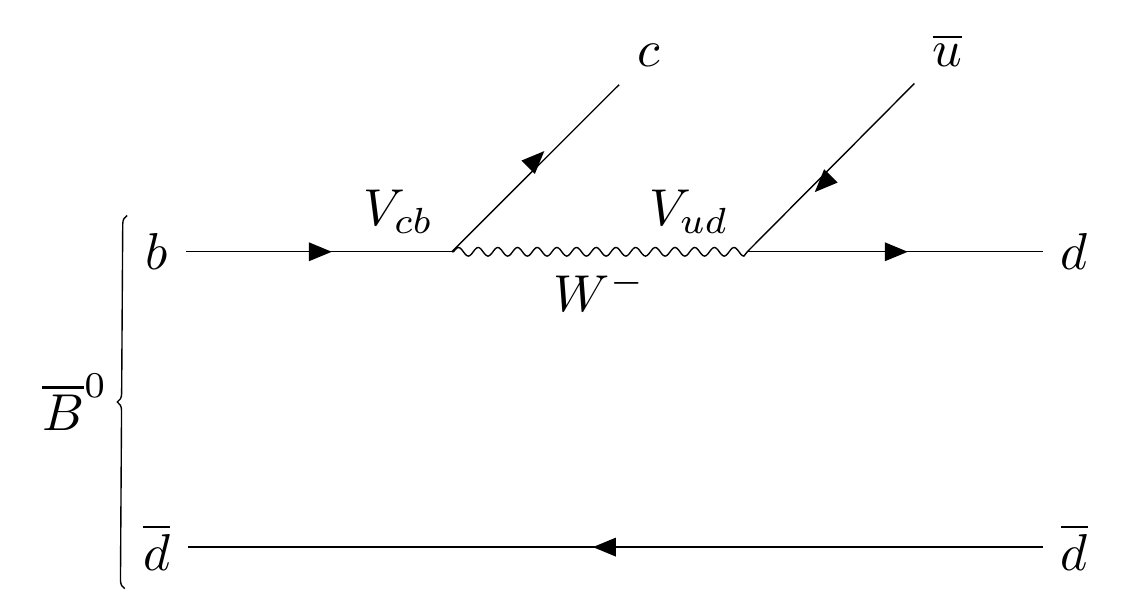}
  \caption{ }
\label{fig:figb}
\end{subfigure}%
\begin{subfigure}{0.49\textwidth}
  \centering
  \includegraphics[width=1\linewidth]{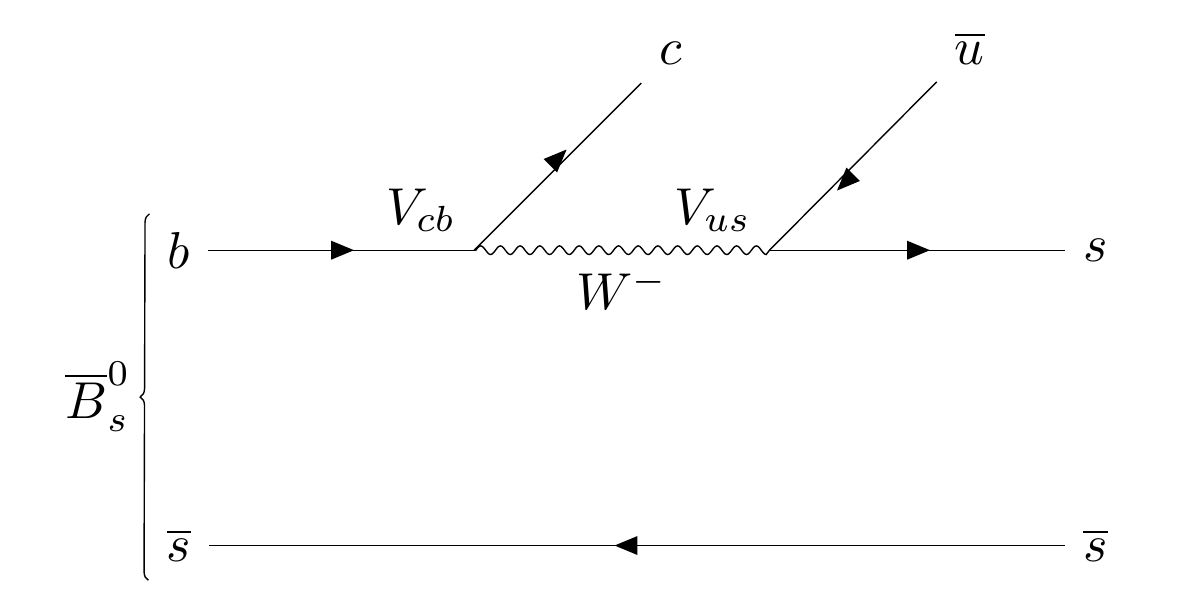}
\caption{ }
\label{fig:fig1a}
\end{subfigure} %
\caption{Feynman diagrams for the decays of $\bar{B}^{0}_{(s)}$ into $D^0$ and a primary $q \bar{q}$ pair. }
\label{fig:fig1}
\end{figure}

For the decays $\bar{B}^0_{(s)} \to D^0 \pi^{+}\pi^{-}$, $\bar{B}^0_{(s)} \to D^0 K^{+}K^{-}$, and $\bar{B}^{0} \rightarrow D^{0} \pi^{0}\eta$, the Feynman diagrams of these decay processes in the quark level are shown in Fig. \ref{fig:fig1}. These processes are Cabibbo suppressed and proceed via the weak transitions of the quarks $b \to s$ and $b \to d$ at the end for the decay of $\bar{B}^0_{s}$ and $\bar{B}^0$, respectively. Considering the fact that the $\bar{d}$ quark in Fig. \ref{fig:figb} and $\bar{s}$ quark in Fig. \ref{fig:fig1a} are spectator, these two processes have similar weak decay mechanism; see more details and discussions in Refs. \cite{Ahmed:2020qkv,Liang:2014tia,Liang:2014ama}.

\subsection{Scalar meson production in $s$-wave}

\begin{figure}
  \centering
  \includegraphics[width=0.6\linewidth]{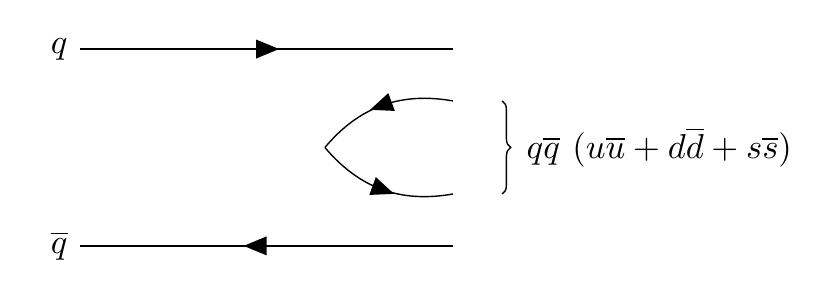}
\caption{Procedure for the hadronization $q\bar{q} \rightarrow q\bar{q}(u\bar{u}+d\bar{d}+s\bar{s})$.}
\label{fig:fig2}
\end{figure}

In this subsection, we discuss the formalism of three pseudoscalar mesons produced in the decays $\bar{B}^0_{(s)} \to D^0 \pi^{+}\pi^{-}$, $\bar{B}^0_{(s)} \to D^0 K^{+}K^{-}$, and $\bar{B}^{0} \rightarrow D^{0} \pi^{0}\eta$ in the quark level, where the final state interaction in the hadron level is taken into account within the ChUA. The dominant weak decay mechanism for the $\bar{B}^{0}_{(s)}$ decays is depicted in Fig. \ref{fig:fig1}, which is proceeded as,
\begin{equation}
\begin{aligned} 
\bar{B}^{0} (b \bar{d}) &\Rightarrow V_{cb} c  \mathit{W^{-}} \bar{d} \Rightarrow V_{cb} V_{ud}^{*} (c\bar{u}) (d\bar{d}) \, ,
\end{aligned} 
\end{equation}
\begin{equation}
\begin{aligned} 
\bar{B}^{0}_{s}(b \bar{s}) &\Rightarrow V_{cb} c \mathit{W^{-}} \bar{s} \Rightarrow V_{cb} V_{us}^{*} (c\bar{u}) (s\bar{s})  \, ,
\end{aligned} 
\end{equation}
where $V_{q_1 q_2}$ is the element of the CKM matrix for the transition of the quarks $q_1 \to q_2$. The $c \bar{u}$ quark pair produces the $D^0$ meson. The other primary $q \bar{q}$ pair creates a pair of mesons $\pi^{+}\pi^{-}$, $\pi^{0}\eta$, or $K^{+}K^{-}$ in the final states. The process of producing two pseudoscalar mesons is proceeded by adding an extra $q \bar{q}$ pairs generated from the vacuum to the primary quark pair, written as $u\bar{u} + d\bar{d} + s\bar{s}$, which is shown in Fig. \ref{fig:fig2}. The hadronization processes are mathematically formulated as
\begin{equation}
\begin{aligned} 
\bar{B}^{0} &\Rightarrow V_{cb} V_{ud}^{*} (c\bar{u} \to  D^0) [d\bar{d} \to d\bar{d} \cdot (u\bar{u}+d\bar{d}+s\bar{s}) ]\, ,
\end{aligned} 
\end{equation}
\begin{equation}
\begin{aligned} 
\bar{B}^{0}_{s} &\Rightarrow V_{cb} V_{us}^{*}  (c\bar{u} \to  D^0)  [ s\bar{s} \to s\bar{s} \cdot (u\bar{u}+d\bar{d}+s\bar{s}) ]\, .
\end{aligned} 
\end{equation}
The procedure of transferring the $q \bar{q}$ pairs to the physical mesons is explicitly expressed as,
\begin{equation}
\begin{aligned}  
d \bar{d} \cdot (u \bar{u}+d \bar{d}+s \bar{s}) & \to (\Phi \cdot \Phi)_{22} =\pi^{+} \pi^{-}+\frac{1}{2} \pi^{0} \pi^{0}-\frac{2}{\sqrt{6}} \pi^{0} \eta+K^{0} \bar{K}^{0}+\frac{1}{3} \eta \eta ,\\ 
s \bar{s} \cdot (u \bar{u}+d \bar{d}+s \bar{s}) & \to (\Phi \cdot \Phi)_{33}=K^{-} K^{+}+K^{0} \bar{K}^{0}+\frac{1}{3} \eta \eta \, ,
\end{aligned}
\label{eq4}
\end{equation} 
where $\Phi$ is the matrix of the meson fields. Taken the standard $\eta - \eta\prime$ mixing into account, the $\Phi$ matrix is written as
\begin{equation}
\Phi=\left(\begin{array}{ccc}{\frac{1}{\sqrt{2}} \pi^{0}+\frac{1}{\sqrt{3}} \eta +\frac{1}{\sqrt{6}} \eta\prime} & {\pi^{+}} & {K^{+}} \\ {\pi^{-}} & {-\frac{1}{\sqrt{2}} \pi^{0}+\frac{1}{\sqrt{3}} \eta +\frac{1}{\sqrt{6}} \eta\prime} & {K^{0}} \\ {K^{-}} & {\bar{K}^{0}} & {-\frac{1}{\sqrt{3}} \eta + \sqrt{\frac{2}{3}} \eta\prime}\end{array}\right).
\label{eq:Mphi}
\end{equation} 

\begin{figure}
  \centering
  \includegraphics[width=1\textwidth]{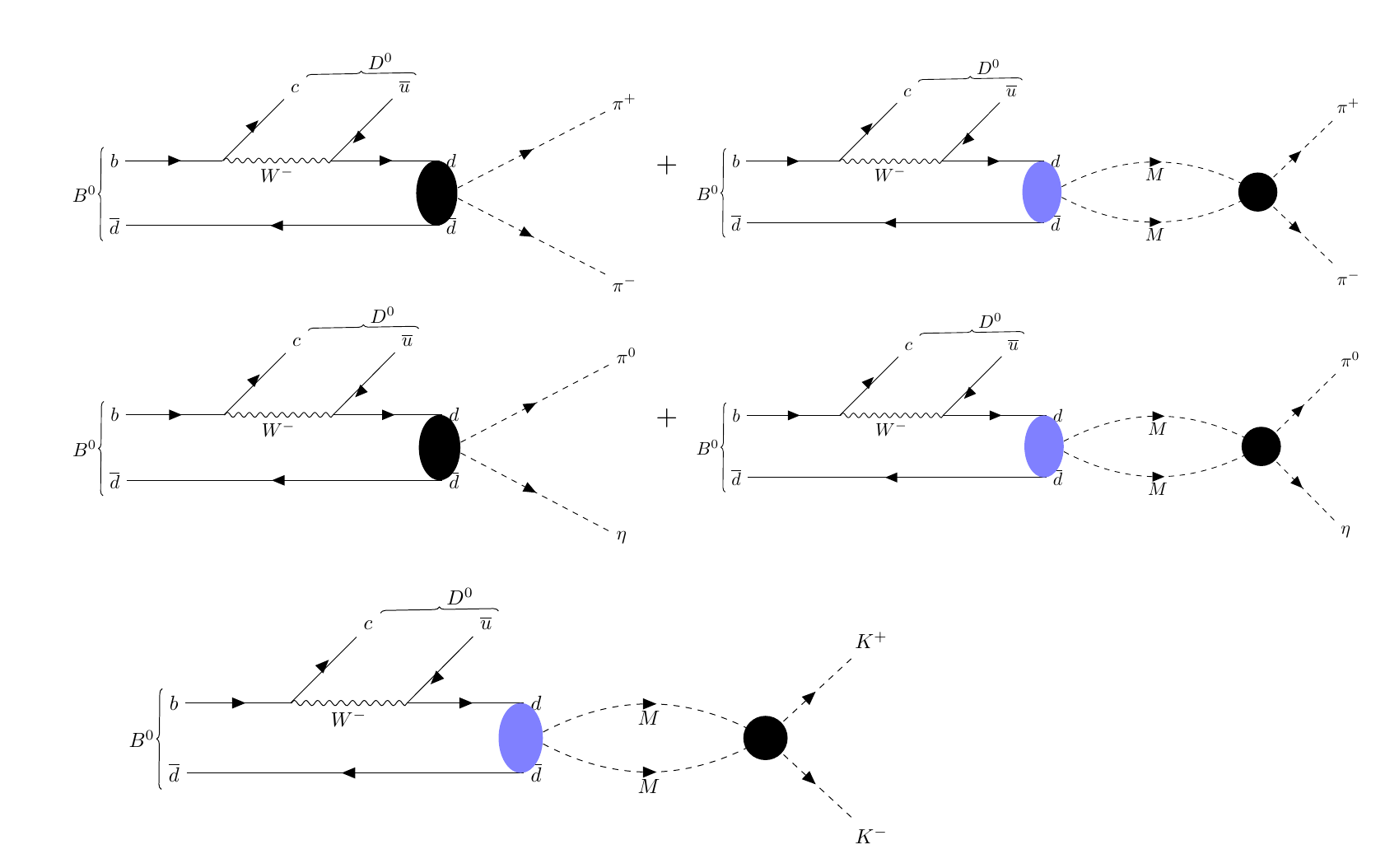}
\caption{Diagrammatic representation of the meson pair production in the final state interactions of the $\bar{B}^{0}$ decay.}
\label{fig:figscattra}
\end{figure} 

\begin{figure}
  \centering
  \includegraphics[width=1\textwidth]{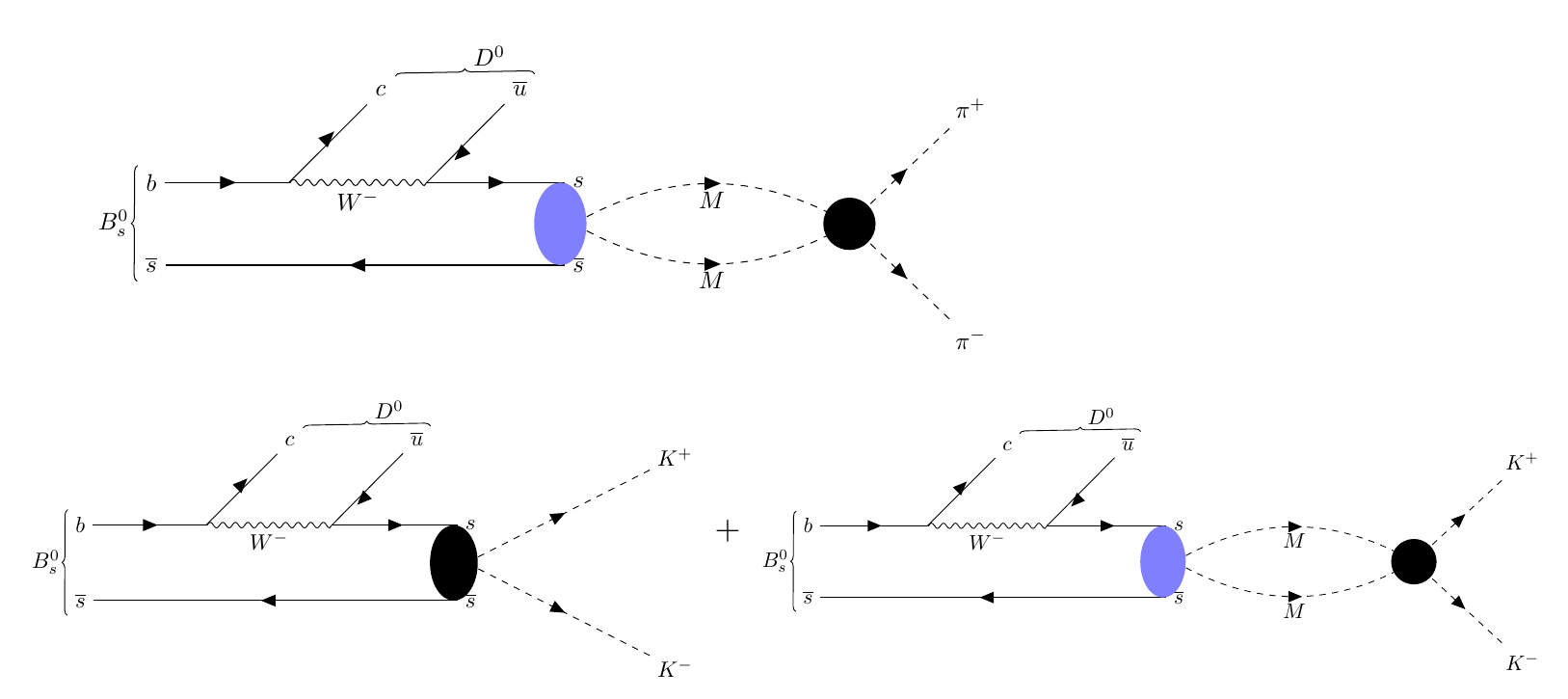}
  \caption{Analogous to Fig. \ref{fig:figscattra}, but for the $\bar{B}^{0}_{s}$ decay.}
\label{fig:figscattrb}
\end{figure}

The final states of Eq. \eqref{eq4} are similar to the one obtained in Ref. \cite{Ahmed:2020qkv}, where only $K\bar{K}$ and $\eta\eta$ channels were produced in the final states of the $\bar{B}^{0}_{s}$ decay. Even though, the final states are hadronized after the weak decay process, they can go to further final state interactions. In the case of the $B^{0}$ decay, we have three pairs of the pseudoscalar mesons, $\pi^{+} \pi^{-}$, $\pi^{0} \eta$, and $K^{+} K^{-}$ alongside the $D^{0}$ meson as depicted in Fig. \ref{fig:figscattra}, where one can see that the  $\pi^{+} \pi^{-}$ and $\pi^{0} \eta$ can be produced from the direct and rescattering mechanisms. However, the $K^{+} K^{-}$ are only produced from the rescattering of the final states. For the $B_{s}^{0}$ decay, the situation is different. The $\pi^{+} \pi^{-}$ are generated from the rescattering of the final states, while the $K^{+} K^{-}$ can be produced from the direct production and the interactions of the final states, as shown in Fig. \ref{fig:figscattrb}. Therefore, the scattering amplitudes of these decay precesses in Figs. \ref{fig:figscattra} and \ref{fig:figscattrb} can be written as
\begin{equation}
\begin{split}
&t_{\bar{B}^{0} \rightarrow D^0 \pi^{+} \pi^{-}} = V_{P} V_{cb} V_{ud}^{*}\left(1 + G_{\pi^{+} \pi^{-}} T_{\pi^{+} \pi^{-} \rightarrow \pi^{+} \pi^{-}} \right.  \\
&\quad+ 2 \frac{1}{2} \frac{1}{2} G_{\pi^{0} \pi^{0}} T_{\pi^{0} \pi^{0} \rightarrow \pi^{+} \pi^{-}} 
\left. +G_{K^{0} \bar{K}^{0}} T_{K^{0} \bar{K}^{0} \rightarrow \pi^{+} \pi^{-}}+ 2 \frac{1}{3} \frac{1}{2} G_{\eta \eta} T_{\eta \eta \rightarrow \pi^{+} \pi^{-}}\right),
\end{split}
\label{B0pipi}
\end{equation}

\begin{equation}
\begin{split}
t_{\bar{B}^{0} \rightarrow D^0 \pi^{0} \eta} =V_{P} V_{cb} V_{ud}^{*}\left(-\frac{2}{\sqrt{6}}-\frac{2}{\sqrt{6}} G_{\pi^{0} \eta} T_{\pi^{0} \eta \rightarrow \pi^{0} \eta} + G_{K^{0} \bar{K}^{0}} T_{K^{0} \bar{K}^{0} \rightarrow \pi^{0} \eta}\right),
\end{split}
\label{B0pieta}
\end{equation}

\begin{equation}
\begin{split}
&t_{\bar{B}^{0} \rightarrow D^0 K^{+} K^{-}} =  V_{P} V_{cb} V_{ud}^{*} \left(G_{\pi^{+} \pi^{-}} T_{\pi^{+} \pi^{-} \rightarrow K^{+} K^{-}} +2 \frac{1}{2} \frac{1}{2} G_{\pi^{0} \pi^{0}} T_{\pi^{0} \pi^{0} \rightarrow K^{+} K^{-}} \right. \\
& \quad +2 \frac{1}{3} \frac{1}{2} G_{\eta \eta} T_{\eta \eta \rightarrow K^{+} K^{-}} 
  -\frac{2}{\sqrt{6}} G_{\pi^{0} \eta} T_{\pi^{0} \eta \rightarrow K^{+} K^{-}}  
  \left. + G_{K^{0} \bar{K}^{0}} T_{K^{0} \bar{K}^{0} \rightarrow K^{+} K^{-}} \right),
\end{split}
\label{B0K+K-}
\end{equation}

\begin{equation}
\begin{split}
&t_{\bar{B}_{s}^{0} \rightarrow D^0 \pi^{+} \pi^{-}} =V_{P} V_{cb} V_{us}^{*} \left(G_{K^{+} K^{-}} T_{K^{+} K^{-} \rightarrow \pi^{+} \pi^{-}} \right.  \\
&\left.\quad +G_{K^{0} \bar{K}^{0}} T_{K^{0}\bar{K}^{0}} \rightarrow \pi^{+} \pi^{-}
+2 \frac{1}{3} \frac{1}{2} G_{\eta \eta} T_{\eta \eta \rightarrow \pi^{+} \pi^{-}}\right),
\end{split}
\label{B0spipi}
\end{equation}

\begin{equation}
\begin{split}
&t_{\bar{B}_{s}^{0} \rightarrow D^0 K^{+} K^{-}}=V_{P} V_{cb} V_{us}^{*} \left(1+ G_{K^{+} K^{-}} T_{K^{+} K^{-} \rightarrow K^{+} K^{-}}\right. \\
&\left.\quad +G_{K^{0} \bar{K}^{0} } T_{K^{0} \bar{K}^{0} \rightarrow K^{+} K^{-}}+2 \frac{1}{3} \frac{1}{2} G_{\eta \eta} T_{\eta \eta \rightarrow K^{+} K^{-}}\right),
\end{split}
\label{B0sK+K-}
\end{equation}
where $V_{P}$ is the production vertex factor, which is universal for these reactions due to the similar weak decay mechanism, and accounts for the direct reproduction diagram. Note that, there is a factor of $2$ in the terms related with the identical particles, such as the $\pi^{0}\pi^{0}$ and $\eta\eta$, which has been cancelled with the factor of $\frac{1}{2}$ in their propagator functions within our normalization scheme. In the case of the $\bar{B}^{0} \rightarrow D^0 K^{+} K^{-}$ decay, we have the contributions from isospins $I = 0$ and $I = 1$, and thus, different isospin contributions in Eq. \eqref{B0K+K-} can be decomposited as,
\begin{equation}
t_{\bar{B}^{0} \rightarrow D^0 K^{+} K^{-}}= t^{I=0} + t^{I=1} \, ,
\end{equation}
where $t^{I=0} $ and $ t^{I=1}$ are given by
\begin{equation}
\begin{aligned}
t^{I=0}=& V_{P} V_{cb} V_{ud}^{*}\left[G_{\pi \pi} T_{\pi^{+} \pi^{-} \rightarrow K^{+} K^{-}}+\frac{1}{2} G_{\pi \pi} T_{\pi^{0} \pi^{0} \rightarrow K^{+} K^{-}}\right. \\
&\left.  +\frac{1}{3} G_{\eta \eta} T_{\eta \eta \rightarrow K^{+} K^{-}} 
+G_{K \bar{K}}\left(\frac{1}{2} T_{K^{0} \bar{K}^{0} \rightarrow K^{+} K^{-}}+\frac{1}{2} T_{K^{+} K^{-} \rightarrow K^{+} K^{-}}\right)\right],
\end{aligned}
\label{tI0}
\end{equation}

\begin{equation}
\begin{aligned}
t^{I=1}=& V_{P} V_{cb} V_{ud}^{*}\left[-\frac{2}{\sqrt{6}} G_{\pi^{0} \eta} T_{\pi^{0} \eta \rightarrow K^{+} K^{-}}\right.\\
&\left.+G_{K \bar{K}}\left(\frac{1}{2} T_{K^{0} \bar{K}^{0} \rightarrow K^{+} K^{-}}-\frac{1}{2} T_{K^{+} K^{-} \rightarrow K^{+} K^{-}}\right)\right].
\end{aligned}
\label{tI1}
\end{equation}

The scattering amplitudes of the coupled channels $T_{ij}$ for the transition of $i \to j$ channel are evaluated from the coupled channel Bethe-Salpeter equation with the on-shell description,
\begin{equation}
T = [1-VG]^{-1}V ,  \label{eq:BS}
\end{equation}
where $V$ matrix is made of the transition potentials. For the coupled channels $\pi^{+}\pi^{-}$, $\pi^{0}\pi^{0}$, $K^{+} K^{-}$, $K^{0} \bar{K}^{0}$, $\eta \eta$, and $\pi^{0} \eta$ with the labels $1$ to $6$, respectively, we take the potentials from Refs. \cite{Oller:1997ti,Xiao:2019lrj},  
\begin{equation}
\begin{aligned}
&V_{11}=-\frac{1}{2 f^{2}} s, \quad V_{12}=-\frac{1}{\sqrt{2} f^{2}}\left(s-m_{\pi}^{2}\right), \quad V_{13}=-\frac{1}{4 f^{2}} s ,\\
&V_{14}=-\frac{1}{4 f^{2}} s, \quad V_{15}=-\frac{1}{3 \sqrt{2} f^{2}} m_{\pi}^{2}, \quad V_{16}=0 ,\\
&V_{22}=-\frac{1}{2 f^{2}} m_{\pi}^{2}, V_{23}=-\frac{1}{4 \sqrt{2} f^{2}} s, \quad V_{24}=-\frac{1}{4 \sqrt{2} f^{2}} s,\\
&V_{25}=-\frac{1}{6 f^{2}} m_{\pi}^{2} , \quad V_{26}=0 , V_{33}=-\frac{1}{2 f^{2}} s, \quad V_{34}=-\frac{1}{4 f^{2}} s ,\\
&V_{35}=-\frac{1}{12 \sqrt{2} f^{2}}\left(9 s-6 m_{\eta}^{2}-2 m_{\pi}^{2}\right),\\
&V_{36}=-\frac{\sqrt{3}}{12 f^{2}} \left(3 s - \frac{8}{3} m_{K}^{2} - \frac{1}{3} m{\pi}^{2}- 6 m_{\eta}^{2}\right) ,\\
&V_{44}=-\frac{1}{2 f^{2}} s , V_{45}=-\frac{1}{12 \sqrt{2} f^{2}}\left(9 s-6 m_{\eta}^{2}-2 m_{\pi}^{2}\right) ,\\
&V_{46}=\frac{\sqrt{3}}{12 f^{2}} \left(3 s - \frac{8}{3} m_{K}^{2} - \frac{1}{3} m{\pi}^{2}- 6 m_{\eta}^{2}\right) ,\\
&V_{55}=-\frac{1}{18 f^{2}}\left(16 m_{K}^{2}-7 m_{\pi}^{2}\right), V_{56}=0 ,\\
&V_{66}= - \frac{1}{3 f^{2}} m_{\pi}^2 \texttt{  .}
\end{aligned}
\end{equation} 

And the diagonal matrix $G$ is constructed by the loop functions of two meson propagators, given by
\begin{equation}
G _ { ii } ( s ) = \int _ { 0 } ^ { q _ { \max } } \frac { q ^ { 2 } d q } { ( 2 \pi ) ^ { 2 } } \frac { \omega _ { 1 } + \omega _ { 2 } } { \omega _ { 1 } \omega _ { 2 } \left[ s - \left( \omega _ { 1 } + \omega _ { 2 } \right) ^ { 2 } + i \varepsilon \right] }   \text{  ,}
\end{equation}
with $q=|\vec{q}\,|$ and $ \omega _ { i } = ( \vec { q } ^ { \:2 } + m _ { i } ^ { 2 } ) ^ { 1 / 2 }$, where the cutoff $q_{max}$ is chosen as $600$ MeV for the case of including $\eta\eta$ channel \cite{Liang:2014tia} and $931$ MeV for the one of excluding $\eta\eta$ channel \cite{Xiao:2019lrj}. 

Finally, the invariant mass distributions for each decay channel can be obtained from the evaluation of the differential decay width $\frac{d\Gamma}{dM_{\text{inv}}}$. Thus, we have
\begin{equation}
\frac{d \Gamma}{d M_{\text{inv}}}=\frac{1}{(2 \pi)^{3}} \frac{1}{4 M_{\bar{B}^0_{(s)}}^{2}} p_{D^0} \tilde{p}_{P}  \bar{\sum} \sum \left| t_{B_{(s)}^{0} \rightarrow D^0 P P^\prime}\right|^{2},
\label{eq10}
\end{equation} 
where $t_{B_{(s)}^{0} \rightarrow D^0 P P^\prime}$ is the scattering amplitude obtained from Eqs. (\ref{B0pipi}-\ref{B0sK+K-}) with $P^{(\prime)}$ stands for a pseudoscalar meson, $p_{D^0}$ the $D^0$ momentum in the rest frame of the decaying $\bar{B}^0_{(s)}$ meson, and $\tilde{p}_{P}$ the kaon (pion) momentum in the rest frame of the $K^{+}K^{-}$ ($\pi^{+}\pi^{-}$, $\pi^{0}\eta$) system, which are given by
\begin{equation}
\begin{aligned}
&p_{D^0}=\frac{\lambda^{1 / 2}\left(M_{\bar{B}^0_{(s)}}^{2}, M_{D^0}^{2}, M_{\text{inv}}^{2}\right)}{2 M_{B_{(s)}}},\\
&\tilde{p}_{P}=\frac{\lambda^{1 / 2}\left(M_{\text{inv}}^{2}, m_{P}^{2}, m_{P^\prime}^{2}\right)}{2 M_{\text {inv }}},
\end{aligned}
\end{equation}
with the usual K\"allen triangle function $\lambda(a, b, c) = a^{2} + b^{2} + c^{2} - 2(ab + ac + bc)$.

\subsection{Vector meson production in $p$-wave} \label{subsec:pwave}

Despite producing the pseudoscalar meson pair alongside $D^{0}$, the vector mesons in $p$-wave also can be generated in these decays, which finally decay into the pseudoscalar meson pair. The $\rho^{0}$ and $\omega$ can be formed from the primary $d\bar{d}$ quarks, and $\phi$ from the $s \bar{s}$ quark pair, which can be decayed into $\pi^+ \pi^-$ and $K^+ K^-$, respectively. As done in Ref. \cite{Liang:2014ama}, we can write down the amplitudes for each channel. It is worth mentioning that in the work of Ref. \cite{Liang:2014ama}, the decay channel of $\bar{B}^{0} \rightarrow D^0 \phi$ had not been considered. The allowed scattering amplitudes in $p$-wave are given by
\begin{equation}
t_{\bar{B}^{0} \rightarrow D^0 \rho^{0}} = \frac{1}{\sqrt{2}} \tilde{V}_{P} V_{cb}  V_{ud}^{*} \textbf{ }, \quad
t_{\bar{B}^{0} \rightarrow D^0 \omega} = \frac{1}{\sqrt{2}} \tilde{V}_{P} V_{cb}  V_{ud}^{*}  \textbf{ },
\end{equation}

\begin{equation}
t_{\bar{B}^{0} \rightarrow D^0 \phi} = \tilde{V}_{P} V_{cb}  V_{ud}^{*}  \textbf{ },  \quad
t_{\bar{B}^{0}_{s} \rightarrow D^0 \phi} = \tilde{V}_{P} V_{cb}  V_{us}^{*} \textbf{ },
\end{equation}
where the prefactor $\frac{-1}{\sqrt{2}}$ and $\frac{1}{\sqrt{2}}$ are the $d\bar{d}$ component in $\rho^{0}$ and $\omega$, respectively, and $\tilde{V}_{P}$ the $p$-wave production vertex factor. The widths for the decays $\bar{B}^0_{(s)} \to D^{0} V$ with a vector meson ($V$) in the products are given by
\begin{equation}
\Gamma_{\bar{B}^{0}_{(s)} \rightarrow D^0 V}=\frac{1}{8 \pi} \frac{1}{m_{B_{(s)}^{0}}^{2}}\left|t_{\bar{B}^{0}_{(s)} \rightarrow D^0 V}\right|^{2} p_{D}.
\label{eq13}
\end{equation} 

Note that, we consider that the resonances $\rho^0$ and $\phi$ can only decay into $\pi^+\pi^-$ and $K^+ K^-$, respectively, of which the contributions to the invariant mass distributions can be obtained by using the spectral function \cite{Liang:2014ama,Wang:2020pem},
\begin{equation}
\frac{d \Gamma_{\bar{B}^{0} \rightarrow D^0 \rho^{0}}}{d M_{\text{inv}}\left(\pi^{+} \pi^{-}\right)}=- \frac{1}{\pi} 2 m_{\rho} \operatorname{Im} \frac{1}{M_{\text{inv}}^{2}-m_{\rho}^{2}+i m_{\rho} \tilde{\Gamma}_{\rho}\left(M_{\text{inv}}\right)} \Gamma_{\bar{B}^{0} \rightarrow D^0 \rho^{0}},
\label{eq14}
\end{equation}

\begin{equation}
\frac{d \Gamma_{\bar{B}^{0}_{s} \rightarrow D^0 \phi }}{d M_{\text{inv}}\left(K^{+} K^{-}\right)}=- \frac{1}{\pi} m_{\phi} \operatorname{Im} \frac{1}{M_{\text{inv}}^{2}-m_{\phi}^{2}+i m_{\phi} \tilde{\Gamma}_{\phi}\left(M_{\text{inv}}\right)} \Gamma_{\bar{B}^{0}_{s} \rightarrow D^0 \phi },
\label{eqphi}
\end{equation}
where $\tilde{\Gamma}_{\rho}\left(M_{\text{inv}}\right)$ and $\tilde{\Gamma}_{\phi}\left(M_{\text{inv}}\right)$ are the energy dependent decay widths of the $\rho^{0}$ decaying into two pions and the $\phi$ decaying into $K^+ K^-$ with $\frac{1}{2}$ weight, respectively, given by
\begin{equation}
\begin{aligned}
&\tilde{\Gamma}_{\rho}\left(M_{\text{inv}}\right)=\Gamma_{\rho}\textbf{ } \left(\frac{p_{\pi}^{\text {off }}}{p_{\pi}^{\text {on }}}\right)^{3},\\
&p_{\pi}^{\mathrm{off}}=\frac{\lambda^{1 / 2}\left(M_{\text{inv}}^{2}, m_{\pi}^{2}, m_{\pi}^{2}\right)}{2 M_{\text{inv}}} \theta\left(M_{\text{inv}}-2 m_{\pi}\right),\\
&p_{\pi}^{\mathrm{on}}=\frac{\lambda^{1 / 2}\left(m_{\rho}^{2}, m_{\pi}^{2}, m_{\pi}^{2}\right)}{2 m_{\rho}},
\end{aligned}
\end{equation}

\begin{equation}
\begin{aligned}
&\tilde{\Gamma}_{\phi}\left(M_{\text{inv}}\right)=\Gamma_{\phi}\textbf{ } \left(\frac{p_{K}^{\text {off }}}{p_{K}^{\text {on }}}\right)^{3},\\
&p_{K}^{\mathrm{off}}=\frac{\lambda^{1 / 2}\left(M_{\text{inv}}^{2}, m_{K}^{2}, m_{\bar{K}}^{2}\right)}{2 M_{\text{inv}}} \theta\left(M_{\text{inv}}-2 m_{K}\right),\\
&p_{K}^{\mathrm{on}}=\frac{\lambda^{1 / 2}\left(m_{\phi}^{2}, m_{K}^{2}, m_{\bar{K}}^{2}\right)}{2 m_{\phi}},
\end{aligned}
\end{equation}
with $p^{\mathrm{on}}(p^{\mathrm{off}})$ the on-shell (off-shell) three momentum in the rest frame.

\section{Results}

In the present work, considering the final state interactions, we look for the resonant contributions in the energy region lower than 1.2 GeV. The coupled channels considered in our work are the ones $\pi^{+}\pi^{-}$, $\pi^{0}\pi^{0}$, $K^{+}K^{-}$, $K^{0}\bar{K}^{0}$, $\eta\eta$, and $\pi^{0} \eta$. As done in the work of \cite{Ahmed:2020qkv}, in our results we also estimate the theoretical uncertainties from including or excluding the $\eta \eta$ channel contribution. In our formalism, the free parameters are the cutoff and the production vertex factors. As discussed above, the value of the cutoff is already taken as $q_{max} = 600$ MeV for the case of including $\eta\eta$ channel \cite{Liang:2014tia} and $931$ MeV for the one of excluding $\eta\eta$ channel \cite{Xiao:2019lrj}. Thus, the only unknown parameters left in our formalism are two production vertex factors, one of which is $V_{p}$ in the $s$-wave case and the other one $\tilde{V_{p}}$ in $p$-wave. More discussions about the partial wave analysis and the final state interactions can be found in the recent review of Ref. \cite{Yao:2020bxx}. Next, we can determine the $V_{p}$ using the experimental branching ratio of the $\bar{B}^{0}\rightarrow D^0 f_{0}(500)$ decay. For the cases of considering the $\eta\eta$ channel and not including the $\eta\eta$ channel, the results  are given by Eq. \eqref{eq:32} and Eq. \eqref{eq:33}, respectively, written
\begin{equation} 
\text{Br}(\bar{B}^{0} \rightarrow D^0 f_{0}(500))= \frac{\Gamma_{\bar{B}^{0} \rightarrow D^0 f_{0}(500)}}{\Gamma_{\bar{B}^{0}}}= \frac{\int_{2m_{\pi}}^{900}\frac{d\Gamma_{\bar{B}^{0} \rightarrow D^0 f_{0}(500)}}{dM_{inv}}dM_{inv}}{\Gamma_{\bar{B}^{0}}}  \\ = \frac{V_{p}^{2}}{\Gamma_{\bar{B}^{0}}} \times (1.92 \times 10^{-5}) \textbf{ },
\label{eq:32}
\end{equation} 
   
 \begin{equation} 
\text{Br}(\bar{B}^{0} \rightarrow D^0 f_{0}(500))= \frac{\Gamma_{\bar{B}^{0} \rightarrow D^0 f_{0}(500)}}{\Gamma_{\bar{B}^{0}}}= \frac{\int_{2m_{\pi}}^{900}\frac{d\Gamma_{\bar{B}^{0} \rightarrow D^0 f_{0}(500)}}{dM_{inv}}dM_{inv}}{\Gamma_{\bar{B}^{0}}}  \\ = \frac{V_{p}^{2}}{\Gamma_{\bar{B}^{0}}} \times (2.58 \times 10^{-5}) \textbf{ }.
\label{eq:33}
\end{equation}
Thus, using the measured branching fraction of $\text{Br}(\bar{B}^{0}\rightarrow D^0 f_{0}(500))= (0.68 \pm 0.08) \times 10^{-4}$ in Particle Data Group \cite{ParticleDataGroup:2020ssz}, we obtain $ \frac{V_{p}^{2}}{\Gamma_{\bar{B}^{0}}}= (3.54 \pm 0.42)$ and $\frac{V_{p}^{2}}{\Gamma_{\bar{B}^{0}}}= (2.63 \pm 0.31)$ for the two cases, respectively, where the uncertainties are estimated from the errors of the experimental branching ratio. Furthermore, for the case of $\tilde{V_{p}}$ in the $p$-wave formalism, using Eq. \eqref{eq13}, we have
\begin{equation}
\text{Br}(\bar{B}^{0} \rightarrow D^0 \rho^{0})= \frac{\Gamma_{\bar{B}^{0} \rightarrow D^0 \rho^{0}}}{\Gamma_{\bar{B}^{0}}}= \frac{\tilde{V}_{P}^{2}}{\Gamma_{\bar{B}^{0}}} \times 0.013.
\end{equation}
Then, taking the experimental results of $\text{Br}(\bar{B}^{0} \rightarrow D^0 \rho^{0})=(3.21 \pm 0.21) \times 10^{-4}$ \cite{ParticleDataGroup:2020ssz}, we can get $\frac{\tilde{V}_{P}^{2}}{\Gamma_{\bar{B}^{0}}}= 0.024 \pm 0.002 $, which is a common value for the $p$-wave calculations.

\begin{figure}
\begin{subfigure}{0.49\textwidth}
  \centering
  \includegraphics[width=1\linewidth]{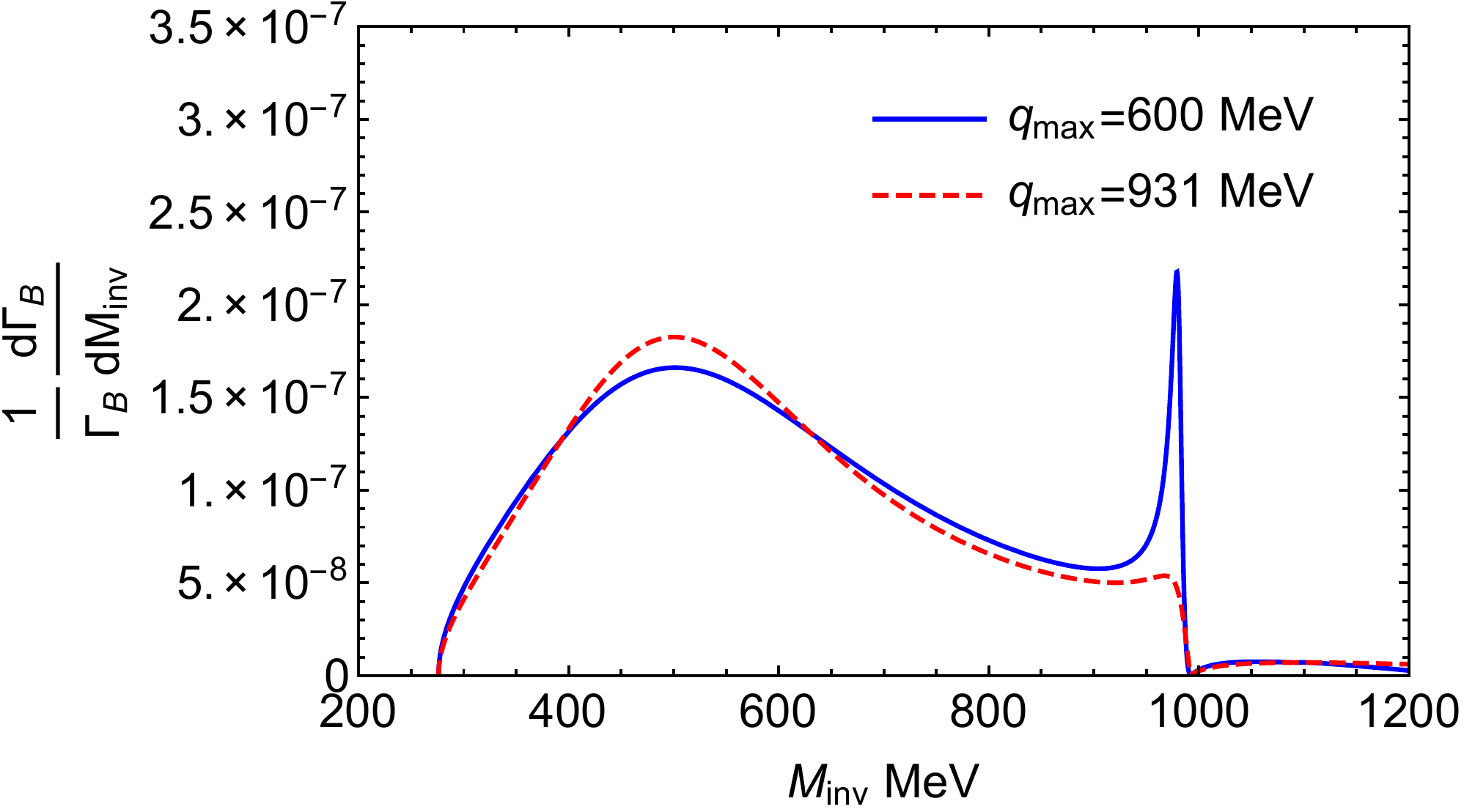}
\caption{For the decay $\bar{B}^{0} \rightarrow D^0 \pi^{+} \pi^{-}$.}
\label{fig:figB0pipi}
\end{subfigure} 
\begin{subfigure}{0.49\textwidth}
  \centering
  \includegraphics[width=1\linewidth]{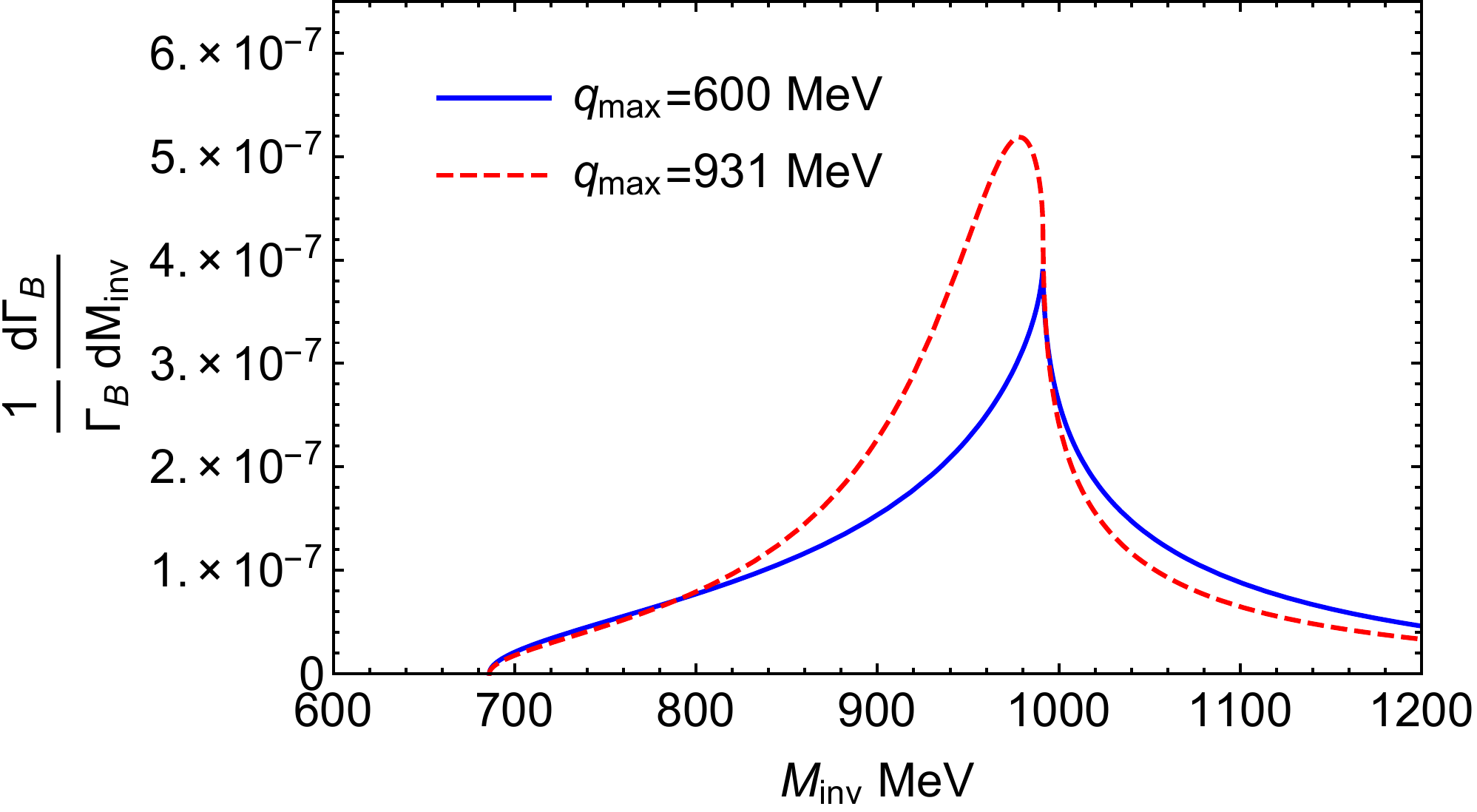}
  \caption{For the decay $\bar{B}^{0} \rightarrow D^0 \pi^{0} \eta$.}
\label{fig:figB0pieta}
\end{subfigure}%
\begin{subfigure}{0.5\textwidth}
  \centering
  \includegraphics[width=1\linewidth]{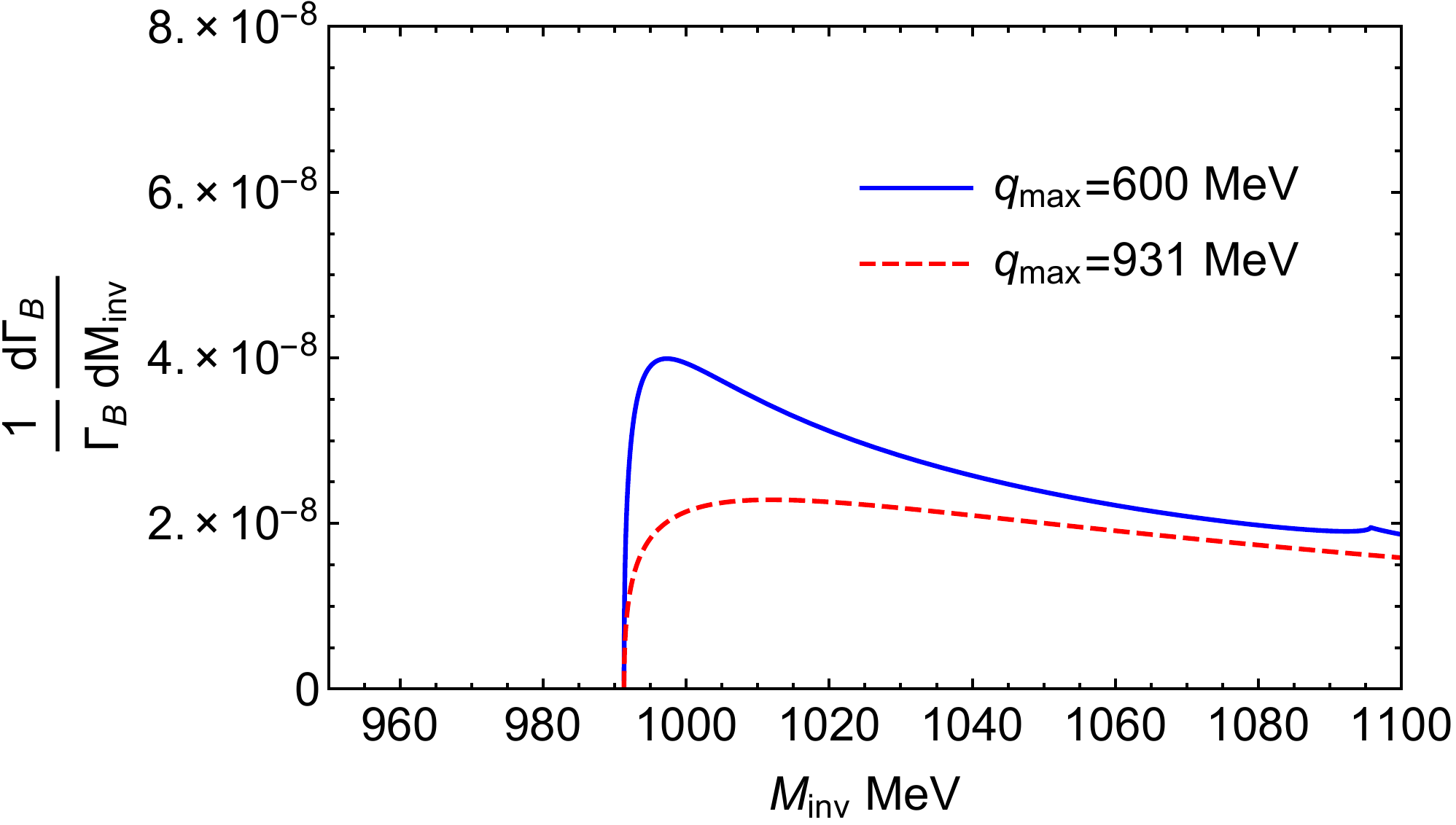}
  \caption{For the decay $\bar{B}^{0} \rightarrow D^0 K^{+} K^{-}$.}
\label{fig:figB0KK}
\end{subfigure}%
\caption{$\pi^+ \pi^-$, $\pi^0 \eta$, and $K^+ K^-$ invariant mass distributions in the decays $\bar{B}^{0} \rightarrow D^0 P P^\prime$, only with $s$-wave contribution. The solid (blue) line is for the results including $\eta \eta$ channel and dashed (red) line for not including it.}
\label{fig:dgammaB0}
\end{figure}

\begin{figure}
  \centering
  \includegraphics[width=0.6\linewidth]{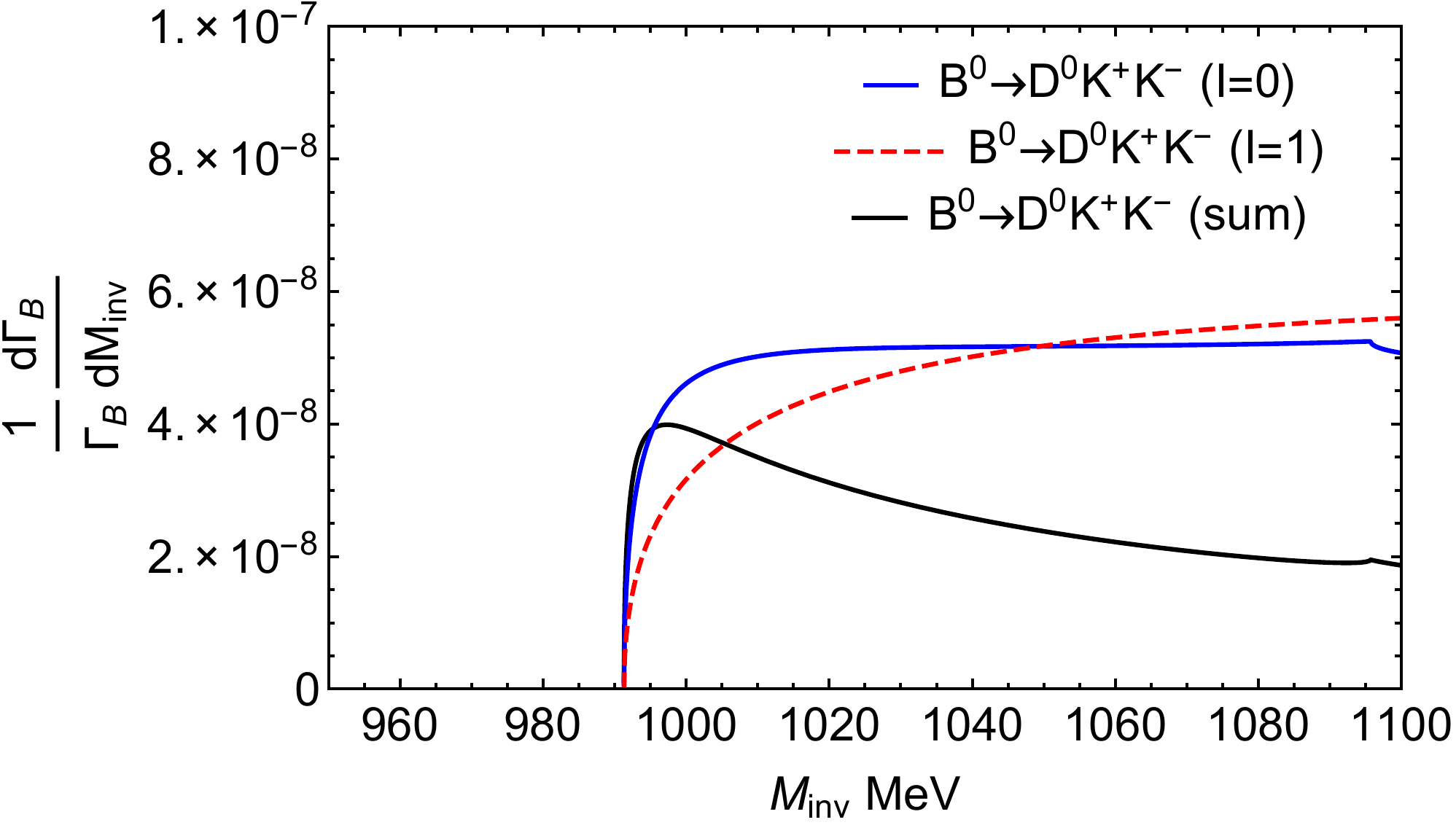}
\caption{Contributions from different isospin components in the $K^+ K^-$ invariant mass distributions in the $\bar{B}^{0} \rightarrow D^0 K^{+} K^{-}$ decay.}
\label{fig:dgammaB0iso}
\end{figure}

\begin{figure}
\begin{subfigure}{0.49\textwidth}
  \centering
  \includegraphics[width=1\linewidth]{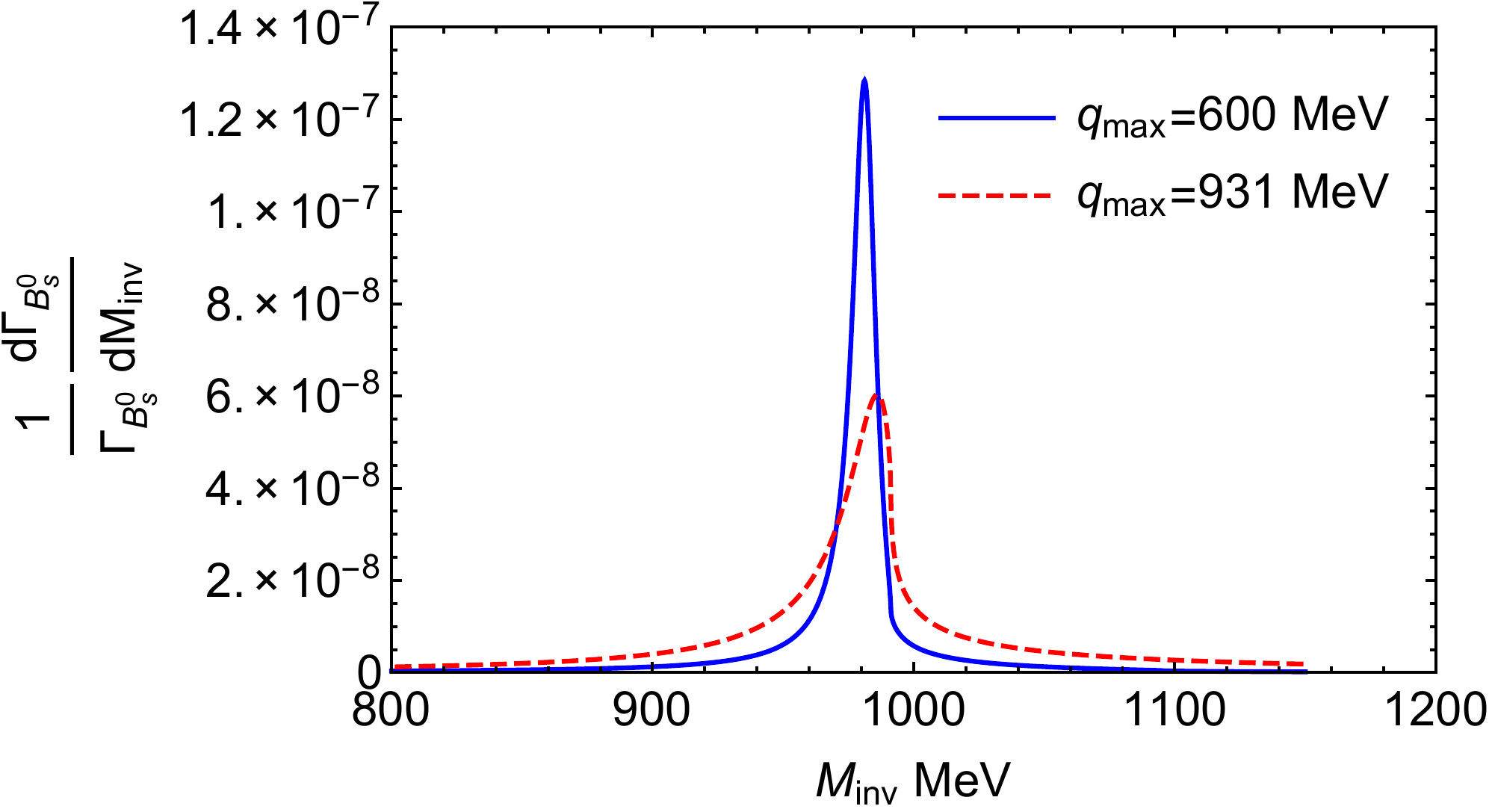}
\caption{For the $\bar{B}_{s}^{0} \rightarrow D^0 \pi^{+} \pi^{-}$ decay.}
\label{fig:figB0spipi}
\end{subfigure} 
\begin{subfigure}{0.49\textwidth}
  \centering
  \includegraphics[width=1\linewidth]{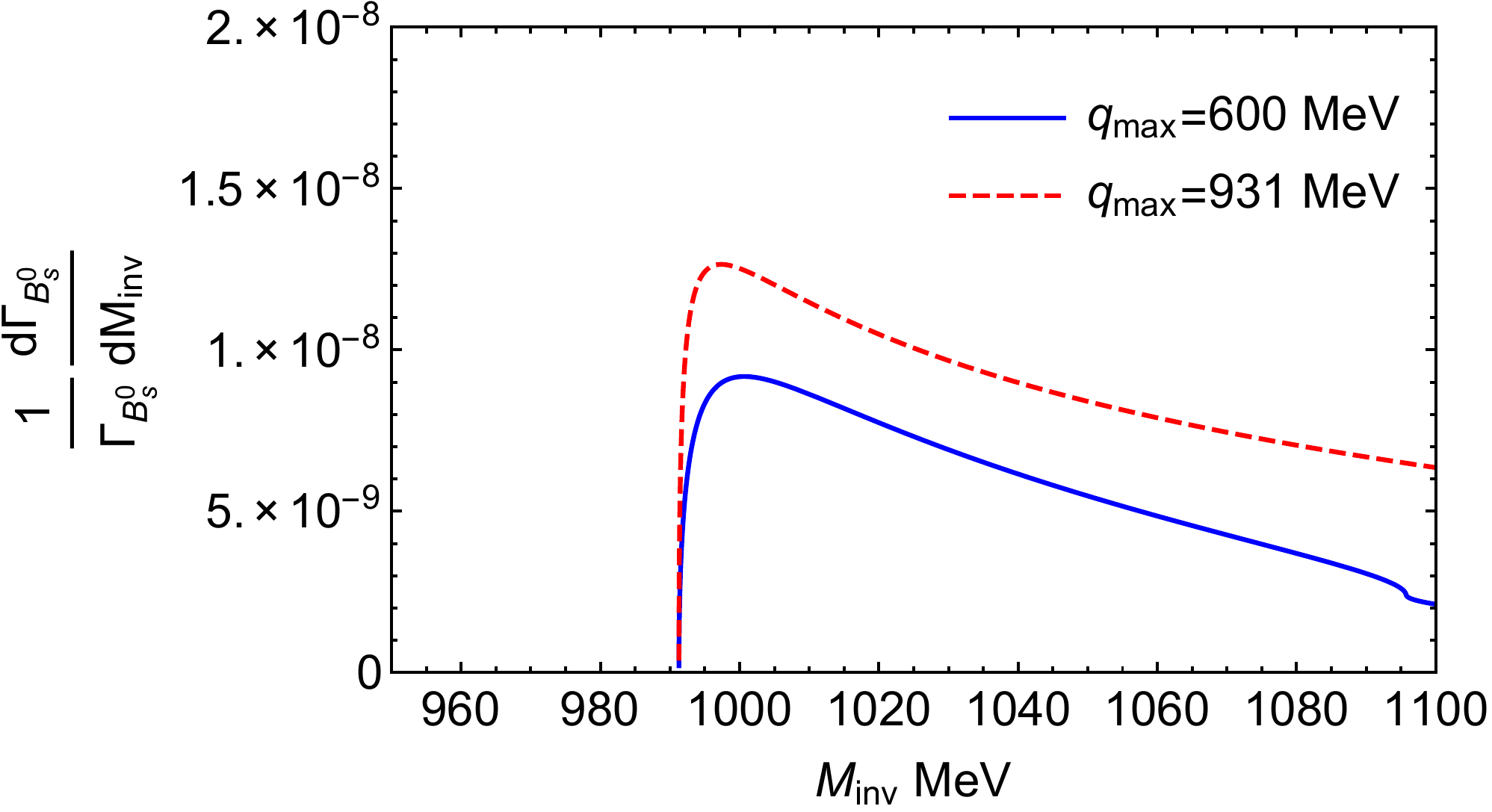}
  \caption{For the $\bar{B}_{s}^{0} \rightarrow D^0 K^{+} K^{-}$ decay.}
\label{fig:figB0sKK}
\end{subfigure}%
\caption{$\pi^+ \pi^-$ and $K^+ K^-$ invariant mass distributions  in the decays $\bar{B}_{s}^{0} \rightarrow D^0 P P^\prime$, only with $s$-wave contributions. The results for different lines are the same as Fig. \ref{fig:dgammaB0}.}
\label{fig:dgammaB0s}
\end{figure}

In Fig. \ref{fig:dgammaB0}, we show the results of the invariant mass distributions of the $\pi^+ \pi^-$, $K^+ K^-$, and $\pi^0 \eta$ channels in the decay processes $\bar{B}^{0} \rightarrow D^0 P P^\prime$, where $P^{(\prime)}$ corresponds to a pseudoscalar meson, which are consistent with the results of Ref. \cite{Liang:2014ama}. Note that in Fig. \ref{fig:dgammaB0} only the $s$-wave contribution is taken into account and the one for $p$-wave discussed later. From the results of Fig. \ref{fig:figB0pipi} for the case of $\bar{B}^{0} \rightarrow D^{0} \pi^+ \pi^-$, one can see that the $f_{0}(500)$ state has a major contribution above the $\pi \pi $ threshold, and the clear signal of the $f_{0}(980)$ resonance is seen especially for the case of including $\eta \eta$ channels, see the solid (blue) line, which shows up as a narrow and small peak near the $K\bar{K}$ threshold. This behaviour of $\pi^+ \pi^-$ mass distribution is similar to the results found in the decays $\bar{B}^{0} \rightarrow J/\psi \pi^{+}\pi^{-}$ \cite{Liang:2014tia} and $\bar{B}^{0} \rightarrow \phi \pi^{+}\pi^{-}$ \cite{Ahmed:2020qkv}. As shown in Fig. \ref{fig:figB0pieta}, the $a_{0}(980)$ state appears in the $\pi^{0} \eta$ mass distributions and has a sizable strength, which is several times stronger than the $f_{0}(980)$ state in Fig. \ref{fig:figB0pipi}, while the normalization factor is the same for both cases. Another feature of Fig. \ref{fig:figB0pieta} is the appearing of cusp effect for the case of including the contribution of $\eta \eta$ channel, where similar effect was found in the work of Ref. \cite{Ahmed:2020kmp}. Note that, for the invariant mass distribution of the $B^{0}$ decaying into $D^{0}$ and two scalar mesons, there would be two contributions from isospins $I=0$ and 1 for the two-body interactions. In the case of $\pi^+ \pi^-$ components, the contribution only comes from the parts of $I=0$, and the case of $\pi^{0} \eta$ channel only has $I=1$ parts contributed. However, the situation is different for the case of $K^{+} K^{-}$ in the final states. In Fig. \ref{fig:figB0KK}, the $K^{+} K^{-}$ invariant mass distribution has a mixing of isospins $I=0$ and 1, which means that this result has the contributions of both $f_{0}(980)$ and $a_{0}(980)$ states. In order to show the strength of each individual state in the $K^{+} K^{-}$ mass distribution, we use the isospin formalism and separate each isospin part as discussed in the formalism, where the results are shown in Fig. \ref{fig:dgammaB0iso}. From Fig. \ref{fig:dgammaB0iso}, one can see that there are interference effects for the sum of two isospin scattering amplitudes for the case of including $\eta \eta$ channel. In Fig. \ref{fig:dgammaB0s}, we show the results of the $\pi^+ \pi^-$ and $K^+ K^-$ invariant mass distributions in the decays $\bar{B}_{s}^{0} \rightarrow D^0 \pi^{+} \pi^{-}$ (Fig. \ref{fig:figB0spipi}) and $\bar{B}_{s}^{0} \rightarrow D^0 K^{+} K^{-}$ (Fig. \ref{fig:figB0sKK}) within the effective energy range of the ChUA. Since each pair of pseudoscalar mesons is produced from the primary $s \bar{s}$ quarks, both of the invariant mass distributions are contributed from the isospin $I=0$ components. As found in the Refs. \cite{Ahmed:2020qkv,Liang:2014tia}, there is no contribution from the $f_{0}(500)$ state for these $\bar{B}_{s}^{0}$ decay channels, and only a large peak in the $\pi^+ \pi^-$ mass distributions, which corresponds to the $f_{0}(980)$ resonance, see Fig. \ref{fig:figB0spipi}. For the case of $K^+ K^-$, the distributions reach the maximum strength close to the $K^+ K^-$ threshold and then fall down gradually, as shown in Fig. \ref{fig:figB0sKK}. As found in Ref. \cite{Ahmed:2020kmp}, the $f_{0}(500)$ state was came from the $\pi^+ \pi^- \to \pi^+ \pi^-$ interaction amplitude only, whereas, the $f_{0}(980)$ state was contributed from the interactions $K^+ K^- \to \pi^+ \pi^-$ and $K^+ K^- \to K^+ K^-$. One can see that there is no interaction part from $\pi^+ \pi^- \to \pi^+ \pi^-$ in Eqs. \eqref{B0spipi} and \eqref{B0sK+K-}. Therefore, it is not surprising that there is no such signal of the $f_{0}(500)$ state in these decays. Besides, one can also see that the results of Fig. \ref{fig:figB0sKK} for the $\bar{B}_{s}^{0} \rightarrow D^0 K^{+} K^{-}$ decay are analogous to the results of Fig. \ref{fig:figB0KK} for the case of $\bar{B}^{0} \rightarrow D^0 K^{+} K^{-}$ decay, even though two precesses have different subprocesses, see Eqs. \eqref{B0K+K-} and \eqref{B0sK+K-} shown in Figs. \ref{fig:figscattra} and \ref{fig:figscattrb}, respectively.

\begin{figure}
  \centering
  \includegraphics[width=0.6\linewidth]{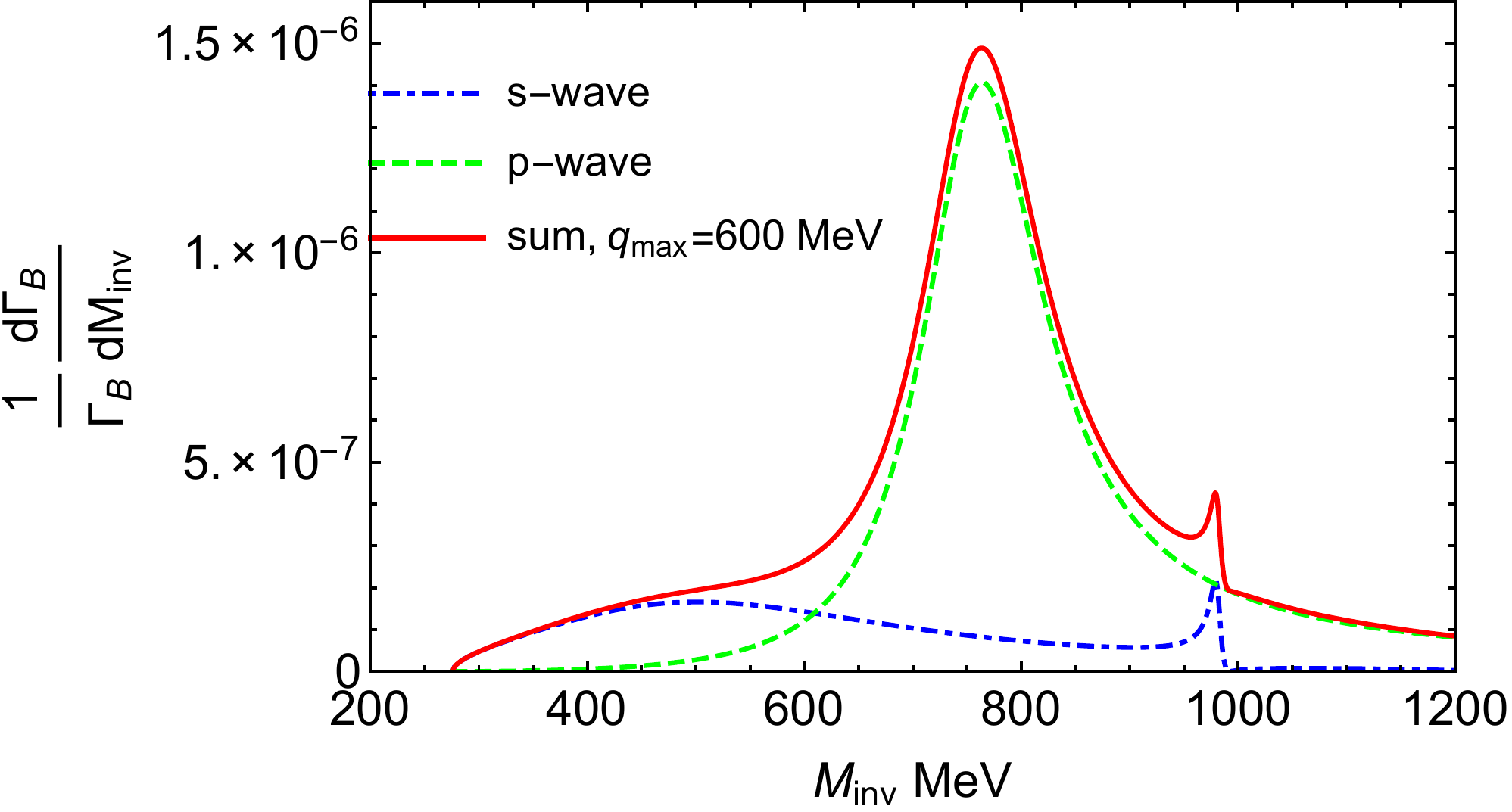}
\caption{$\pi^+ \pi^-$ invariant mass distributions in the $\bar{B}^{0} \rightarrow D^0 \pi^{+} \pi^{-}$ decay, where the contributions from the $s$-wave and $p$-wave are shown as dashed-dot (blue) line and dashed (green) line, respectively, and the sum of two of them is displayed as solid (red) line.}
\label{fig:dgammaB0sum}
\end{figure}

\begin{figure}
  \centering
  \includegraphics[width=0.6\linewidth]{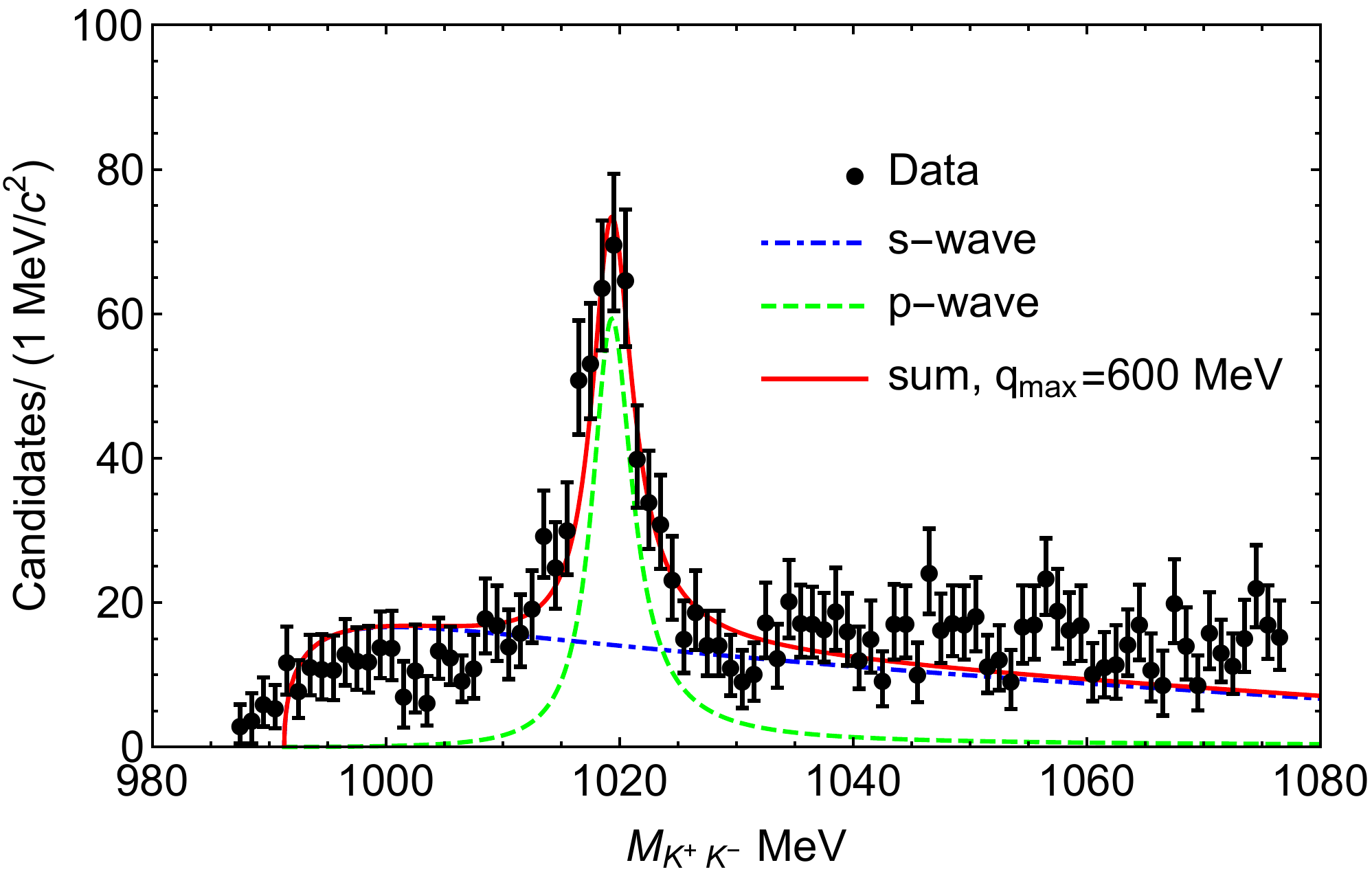}
\caption{$K^+ K^-$ invariant mass distributions in the $\bar{B}_{s}^{0} \rightarrow D^0 K^{+} K^{-}$ decay, where the contributions from the $s$-wave and $p$-wave are shown as dashed-dot (blue) line and dashed (green) line, respectively. The red line represents the sum of the $s$- and $p$-waves contributions (with $\chi^{2}/dof = 1.63$), and the black dots are the experimental data \cite{Aaij:2018jqv}.}
\label{fig:dgammaB0phi}
\end{figure}

The results of Figs. \ref{fig:dgammaB0} and \ref{fig:dgammaB0s} are only taken into account the $s$-wave contribution. However, as discussed in the subsection \ref{subsec:pwave} the $p$-wave also has a significant contribution for the decays channels $\pi^+ \pi^-$ and $K^+ K^-$, since the vector mesons $\rho$ and $\omega$ can decay into the $\pi^+ \pi^-$ channel and the $\phi$ into the $K^+ K^-$. The results of considering the $s$-wave and $p$-wave contributions are shown in Figs. \ref{fig:dgammaB0sum} and \ref{fig:dgammaB0phi} for the decays $\bar{B}^{0} \rightarrow D^0 \pi^{+} \pi^{-}$ and $\bar{B}_{s}^{0} \rightarrow D^0 K^{+} K^{-}$, respectively. In Fig. \ref{fig:dgammaB0sum}, one can see that the $p$-wave contribution of the $\rho$ meson is dominated the invariant mass distributions, which are stronger than the contributions from the $f_{0}(500)$ and $f_{0}(980)$ states in $s$-wave, and also consistent with the results of Ref. \cite{Liang:2014ama}. Indeed, this coincides with the fact that the branching ratio for the decay $\bar{B}^{0} \rightarrow D^0 \rho^{0}$ is bigger than the result for the decay $\bar{B}^{0} \rightarrow D^0 f_{0}(500)$ \cite{ParticleDataGroup:2020ssz}; see the discussions above for determining the production vertex factors. Even though the contributions from the $f_{0}(500)$ and $f_{0}(980)$ states are small, their signals can be clear seen in the line shape of the $\pi^+ \pi^-$ invariant mass distributions.
Likewise, the contribution of the $\phi$ meson in $p$-wave surpasses the ones of the $f_{0}(980)$ for the $K^+ K^-$ invariant mass distributions in the case of the $\bar{B}_{s}^{0} \rightarrow D^0 K^{+} K^{-}$ decay; see the results of Fig. \ref{fig:dgammaB0phi}. Note that, in Fig. \ref{fig:dgammaB0phi} we have two arbitrary normalization factors to match the experimental events, one for the $s$-wave part and the other one for the $p$-wave contribution. These two factors are determined from the fitting of the experimental data \cite{Aaij:2018jqv} with the sum of the contributions from the $s$- and $p$-waves, of which the values are $1.81$ and $2.47$ for the $s$- and $p$-waves, respectively, with $\chi^{2}/dof = 1.63$. One can see that our fitting results describe the experimental data well with the contributions from both the $s$- and $p$-waves. It looks like the signal of $f_{0}(980)$ resonance is not so visible. Once again, the results of Fig. \ref{fig:dgammaB0phi} show that the dominant contribution comes from the vector meson primarily produced in $p$-waves, which indeed indicate the molecular nature of the resonances $f_{0}(500)$, $f_{0}(980)$ and $a_{0}(980)$ generated from the subprocesses of final state interactions and not produced in the primary $q\bar{q}$ quark pair, as found in Ref. \cite{Wang:2021nxz}. Analogous considerations can be found in Ref. \cite{Hyodo:2008xr} for the origin nature of the resonances.

Next, we make a further calculation of the branching fractions for each decay channel. The obtained branching fractions and the ratios of branching fractions are given in Tables \ref{tab:tab1} and \ref{tab:tab2}. From the results in Table \ref{tab:tab1}, one can see the consistency between our work and the experimental measurements within the uncertainties. The uncertainties in our model come from the errors of the branching fractions used to determine the vertex factors, as discussed above. In Table \ref{tab:tab1}, for the case of including $\eta \eta$ channel, the branching fraction of $\bar{B}^{0}\rightarrow D^0 f_{0}(980)$ is in good agreement with the experiment, whereas, for the one of eliminating $\eta \eta$ channel, the result is affected to give a bit small value. This is analogous to the case of the $\bar{B}^{0} \rightarrow \phi \pi^{+}\pi^{-}$ decay found in Ref. \cite{Ahmed:2020qkv}. Besides, for the case of $\bar{B}^{0}\rightarrow D^0 \pi^+ \pi^-$ decay, if we consider full decay mode, including or excluding $\eta \eta$ channel does not influence the outcomes; see the second row of Table \ref{tab:tab1}. Moreover, the predicted branching ratio of $\bar{B}^{0}\rightarrow D^0 a_{0}(980)$ decay is one magnitude larger than the result of $\bar{B}^{0}\rightarrow D^0 f_{0}(980)$ decay, which is easily expected from the results of Fig. \ref{fig:dgammaB0}. The results of $\text{Br}(\bar{B}^{0}_{s}\rightarrow D^0 f_{0}(980))$ and $\text{Br}(\bar{B}^{0}_{s}\rightarrow D^0 K^+ K^-)$ are also consistent with the upper limit of the experimental measurements for the cases of both including and excluding the $\eta \eta$ channel. Moreover, with the branching fractions obtained, one can easily evaluate their ratios, as shown in Table \ref{tab:tab2}, some of which are consistent with the experimental results within the uncertainties. Note that, these values are not depended on the production vertex factor, since it is universal for all the calculations with the contributions from $s$-wave and cancelled in these fractions. Therefore, these ratios are more reliable in our formalism. 

\begin{table}
\renewcommand{\arraystretch}{1.7}
     \setlength{\tabcolsep}{0.2cm}
\center
\caption{Results of the branching fractions.}
\resizebox{0.8\textwidth}{!}{\begin{tabular}{|c|c|c|c|}
\hline
Branching ratios & Without $\eta\eta$ channel  & With $\eta\eta$ channel  & Exp. \cite{ParticleDataGroup:2020ssz}  \\ \hline 
$\text{Br}(\bar{B}^{0}\rightarrow D^0 f_{0}(980))$  &  $(5.47 \pm 0.64 ) \times 10^{-6} $   &  $(8.02 \pm 0.94 ) \times 10^{-6} $   &  $8.0 \pm 4.0 \times 10^{-6} $  \\ \hline  
$\text{Br}(\bar{B}^{0}\rightarrow D^0 \pi^+ \pi^-)$  &    $(7.15 \pm 0.51)\times 10^{-4}$      & $(7.17 \pm 0.51)\times 10^{-4}$  & $(8.8 \pm 0.5) \times 10^{-4}$  \\ \hline
$\text{Br}(\bar{B}^{0}\rightarrow D^0 a_{0}(980)$  &    $(7.34 \pm 0.86)\times 10^{-5}$      & $(6.08 \pm 0.72)\times 10^{-5}$  & $\cdots$  \\ \hline
$\text{Br}(\bar{B}^{0}\rightarrow D^0 K^+ K^-)$  &    $(6.96 \pm 0.47)\times 10^{-5}$      & $(7.07 \pm 0.49)\times 10^{-5}$  & $(6.1 \pm 0.4 \pm 0.3 \pm 0.3) \times 10^{-5} $  \\ \hline
$\text{Br}(\bar{B}^{0}_{s}\rightarrow D^0 f_{0}(980))$  &    $(3.10 \pm 0.37)\times 10^{-6}$      & $(2.16 \pm 0.25)\times 10^{-6}$  & $< 3.1 \times 10^{-6}$  \\ \hline
$\text{Br}(\bar{B}^{0}_{s}\rightarrow D^0 K^+ K^-)$  &    $(3.41 \pm 0.23)\times 10^{-5}$      & $(3.35 \pm 0.22)\times 10^{-5}$  & $(5.7 \pm 0.5 \pm 0.4 \pm 0.5) \times 10^{-5}$  \\ \hline
\end{tabular}}
\label{tab:tab1}
\end{table}

\begin{table}
\renewcommand{\arraystretch}{1.7}
     \setlength{\tabcolsep}{0.2cm}
\center
\caption{Ratios of the branching fractions.}
\resizebox{0.75\textwidth}{!}{\begin{tabular}{|c|c|c|c|}
\hline
   Ratios & Without $\eta\eta$ channel  & With $\eta\eta$ channel  & Exp. \cite{ParticleDataGroup:2020ssz}  \\ \hline 
   $\frac{\text{Br}(\bar{B}^{0}\rightarrow D^0 K^+ K^-)}{\text{Br}(\bar{B}^{0}\rightarrow D^0 \pi^+ \pi^-)}$  &    $0.097 \pm 0.014$     & $0.098 \pm 0.014$  & $0.069 \pm 0.004 \pm 0.003$ \\ \hline   
   $\frac{\text{Br}(\bar{B}^{0}_{s}\rightarrow D^0 K^+ K^-)}{\text{Br}(\bar{B}^{0}\rightarrow D^0 K^+ K^-)}$  &    $0.49 \pm 0.07$      & $0.47 \pm 0.06$  & $0.93 \pm 0.089 \pm 0.069$ \\ \hline 
   $\frac{\text{Br}(\bar{B}^{0}\rightarrow D^0 f_{0}(980))}{\text{Br}(\bar{B}^{0}\rightarrow D^0 f_{0}(500))}$  &    $0.08 \pm 0.02$    & $0.12 \pm 0.03$   & $0.12 \pm 0.06$\\ \hline 
    $\frac{\text{Br}(\bar{B}^{0}\rightarrow D^0 f_{0}(980))}{\text{Br}(\bar{B}^{0}\rightarrow D^0 a_{0}(980))}$  &    $0.07 \pm 0.02$    & $0.13 \pm 0.03$  & $\cdots$ \\ \hline 
\end{tabular}}
\label{tab:tab2}
\end{table}

Furthermore, we can obtain the ratios of branching fractions and the branching fractions for the other vector decay channels $D^{0} V$. For these cases, the results are related to the CKM matrix elements for the intermediate weak decay procedures, and thus, one can easily get the ratios as below, 
\begin{equation}\begin{array}{l}
R_{1}^{th}=\frac{\Gamma_{\bar{B}^{0} \rightarrow D^{0} \rho^{0}}}{\Gamma_{\bar{B}^{0}_{s} \rightarrow D^{0} \phi}}=\frac{1}{2}\left|\frac{V_{cb}  V_{u d}^{*}}{V_{cb}  V_{u s}^{*} }\right|^{2} \frac{m_{\bar{B}^{0}_{s}}^{2}}{m_{\bar{B}^{0}}^{2}} (\frac{p_{\rho^{0}}}{p_{\phi}})^{3}= 9.82,  \\
R_{2}^{th}=\frac{\Gamma_{\bar{B}^{0} \rightarrow D^{0} \rho^{0}}}{\Gamma_{\bar{B}^{0} \rightarrow D^{0} \omega}}=(\frac{p_{\rho^{0}}}{p_{\omega}})^{3}= 1.002 .
\label{eq:ratio1}
\end{array}\end{equation}
Besides, using the measured branching ratio of $\bar{B}^{0} \rightarrow D^{0} \rho^{0}$ decay to determine the production vertex factor as discussed above, we can also obtain the other two branching ratios,
\begin{equation}\begin{array}{l}
\text{Br}(\bar{B}^{0}_{s} \rightarrow D^{0} \phi) = \frac{\Gamma_{\bar{B}^{0}_{s} \rightarrow D^{0} \phi}}{\Gamma_{B}}= (3.27 \pm 0.21) \times 10^{-5},\\
\text{Br}(\bar{B}^{0} \rightarrow D^{0} \omega)=\frac{\Gamma_{\bar{B}^{0} \rightarrow D^{0} \omega}}{\Gamma_{B}}= (3.2 \pm 0.21) \times 10^{-4}.
\label{vectorBr}
\end{array}\end{equation}
which are consistent with the experimental results \cite{ParticleDataGroup:2020ssz} within the uncertainties,
\begin{equation}\begin{array}{l}
\text{Br}(\bar{B}^{0}_{s} \rightarrow D^{0} \phi) = (3.0 \pm 0.5) \times 10^{-5},\\
\text{Br}(\bar{B}^{0} \rightarrow D^{0} \omega)= (2.54 \pm 0.16) \times 10^{-4}.
\end{array}\end{equation}

\section{Conclusions}

In the present work, based on the chiral unitary approach, we utilize the final state interaction formalism to investigate the decays $\bar{B}^0_{(s)} \to D^0 \pi^{+}\pi^{-}$, $\bar{B}^0_{(s)} \to D^0 K^{+}K^{-}$, and $\bar{B}^{0} \rightarrow D^{0} \pi^{0}\eta$. Moreover, we take into account the resonant contributions from both the $s$- and $p$-waves in the relevant decay channels. The considered resonance contributions come from the states $f_{0}(500)$, $f_{0}(980)$, $\rho$, and $\phi$ in the $s$- and $p$-wave calculations for the decays $\bar{B}^0 \to D^0 \pi^{+}\pi^{-}$ and $\bar{B}^0 \to D^0 K^{+}K^{-}$, the state $a_{0}(980)$ for the $\bar{B}^{0} \rightarrow D^{0} \pi^{0}\eta$ decay, and the states $f_{0}(980)$ and $\phi$ in the $s$- and $p$-wave contributions for the decays $\bar{B}^0_{s} \to D^0 \pi^{+}\pi^{-}$ and $\bar{B}^0_{s} \to D^0 K^{+}K^{-}$. From the invariant mass distribution of $B^{0}$ decays, it is found that the $f_{0}(500)$ and $a_{0}(980)$ has sizeable structures compared to the $f_{0}(980)$ state, which leads to the fact that the decaying into $f_{0}(980)$ state has a small branching fraction. However, for the case of $B^{0}_{s}$ decays, since the hadronization in these decay processes come from the primary quark pair $s\bar{s}$, there is no contribution from isospin $I=1$. Thus, there is only a peak closed to the $K^{+} K^{-}$ threshold, which corresponds to the state $f_{0}(980)$, and no signal for the resonance $f_{0}(500)$. Despite the $s$-wave contributions are considered, from Figs. \ref{fig:dgammaB0sum} and \ref{fig:dgammaB0phi} one can see that  the vector mesons $\rho$ and $\phi$, produced in the primary $q\bar{q}$ quark pair from the $p$-wave contributions, have the dominant contributions in the decays $\bar{B}^0 \to D^0 \pi^{+}\pi^{-}$ and $\bar{B}^0_{s} \to D^0 K^{+}K^{-}$, respectively, which happen to indicate the molecular nature of the resonances $f_{0}(500)$, $f_{0}(980)$ and $a_{0}(980)$ generated from the subprocesses of final state interactions. In addition to show the invariant mass distributions, one can also calculate the branching fractions and the ratios of branching fractions. We investigate the branching ratios for different decay processes with the scalar and vector mesons produced in the final states, where some of our results are in agreement with the experiments. We also predict the branching fraction of $\bar{B}^{0} \rightarrow D^{0} a_{0}(980) [a_{0}(980) \to \pi^{0}\eta]$ decay, obtained as $\text{Br}(\bar{B}^{0}\rightarrow D^0 a_{0}(980)=(7.34 \pm 0.86)\times 10^{-5}$ and $\text{Br}(\bar{B}^{0}\rightarrow D^0 a_{0}(980)=(6.08 \pm 0.72)\times 10^{-5}$ for the cases of without and with $\eta \eta$ channel contribution, respectively. 
Then, the relative weights of the branching fractions are predicted with no free parameters due to the cancellation of the production vertex factor. And thus, these predictions are meaningful for future experiments, such as the ratio of $\text{Br}(\bar{B}^{0}\rightarrow D^0 f_{0}(980))/\text{Br}(\bar{B}^{0}\rightarrow D^0 a_{0}(980))$.

\section*{Acknowledgments}

We thank Prof. Mei Huang for careful reading the manuscript and useful comments, and acknowledge Profs. Eulogio Oset and Nikolay Achasov for helpful comments.

 \addcontentsline{toc}{section}{References}
\end{document}